\documentclass[10pt,a4paper]{article}

\usepackage[utf8]{inputenc}
\usepackage[T1]{fontenc}
\usepackage{lmodern}
\usepackage{microtype}
\usepackage[margin=2.3cm,top=2.2cm,bottom=2.4cm]{geometry}
\usepackage{amsmath,amssymb,amsthm}
\usepackage{booktabs}
\usepackage{enumitem}
\usepackage{tikz}
\usetikzlibrary{decorations.pathreplacing,arrows.meta,positioning,fit,backgrounds,calc}
\usepackage[hidelinks]{hyperref}
\usepackage{placeins}
\hypersetup{pdftitle={Finding Integral Stable Matchings is PPAD-Complete: Unimodular Hypergraphs, Network b-Matchings, and Kernels in Perfect Graphs},pdfauthor={Gergely Cs\'aji}}
\newtheorem{theorem}{Theorem}
\newtheorem{lemma}[theorem]{Lemma}
\newtheorem{corollary}[theorem]{Corollary}
\newtheorem{observation}[theorem]{Observation}
\newtheorem{proposition}[theorem]{Proposition}
\theoremstyle{definition}
\newtheorem{definition}[theorem]{Definition}
\theoremstyle{remark}

\DeclareRobustCommand{\NOR}{\textnormal{\textsc{nor}}}
\DeclareRobustCommand{\PUR}{\textnormal{\textsc{purify}}}
\DeclareRobustCommand{\PC}{\textnormal{\textsc{Pure-Circuit}}}
\DeclareRobustCommand{\UDA}{\textnormal{UDA}}
\DeclareRobustCommand{\SHM}{\textnormal{SHM}}
\DeclareRobustCommand{\SHbM}{\textnormal{SH$b$M}}
\DeclareRobustCommand{\HRC}{\textnormal{HRC}}
\DeclareRobustCommand{\NFHRC}{\textnormal{Near-Feasible HRC}}
\DeclareRobustCommand{\COPY}{\textnormal{\textsc{copy}}}
\DeclareRobustCommand{\OR}{\textnormal{\textsc{or}}}
\DeclareRobustCommand{\AND}{\textnormal{\textsc{and}}}
\DeclareRobustCommand{\NOT}{\textnormal{\textsc{not}}}
\DeclareRobustCommand{\BRK}[1]{\textnormal{\textsc{break}}\ensuremath{_{#1}}}
\newcommand{\PPAD}{\textsf{PPAD}}
\newcommand{\FP}{\textsf{FP}}
\newcommand{\doi}[1]{\href{https://doi.org/#1}{\nolinkurl{#1}}}
\newcommand{\home}{\mathsf{h}}
\newcommand{\inp}{\mathsf{in}}

\title{Finding Integral Stable Matchings is PPAD-Complete:\\ Unimodular Hypergraphs, Network $b$-Matchings,\\ and Kernels in Perfect Graphs}
\author{Gergely Cs\'aji\\[2pt]
\small HUN-REN SZTAKI and Budapest University of Technology and Economics\\
\small \texttt{csajigergely@sztaki.hu}}
\date{}

\begin{document}
\maketitle

\begin{abstract}
Polyhedral integrality guarantees integral stable solutions in several classical stability problems: stable matchings in normal hypergraphs, stable $b$-matchings in unimodular hypergraphs, and kernels in clique-acyclic orientations of perfect graphs. Whether they can be computed efficiently has been a long-standing open question; for kernels in perfect graphs it has been open since the existence theorem of Boros and Gurvich in 1996. We show that, unless $\PPAD\subseteq\FP$, they cannot. Finding a stable matching is \PPAD-complete already in linear unimodular hypergraphs with hyperedges of size at most three and vertex degrees at most three. Consequently, finding a kernel in a clique-acyclic orientation of a perfect graph is \PPAD-hard even if all cliques have at most three vertices, answering an open question of Pass-Lanneau, Igarashi and Meunier, and stable matchings are \PPAD-hard to find in balanced and in normal hypergraphs. This contrasts with the polynomial algorithm of Chandrasekaran and Kalathur for network hypergraphs. With capacities, even network hypergraphs are hard: University Dual Admission is \PPAD-complete with capacities at most nine, so stable $b$-matching is \PPAD-complete on network hypergraphs with hyperedges of size three. This yields \PPAD-completeness for a natural class of hospitals/residents markets with couples where stable matchings always exist, and \PPAD-hardness of computing the near-feasible stable matchings, where capacities are flexible. Both reductions are from \PC{} with constant-size gadgets. For unit capacities we use a dual-rail encoding, a symmetry-breaking gadget made of two coupled $2\times2$ stable marriage markets, and a parity criterion on the bipartite incidence graph that certifies total unimodularity. We expect these tools to be useful for further hardness results about problems with guaranteed integral solutions.
\end{abstract}

\section{Introduction}

In the classical bipartite model, a stable matching always exists and can be found by the deferred acceptance algorithm~\cite{GS62}. In stable hypergraph matching (\SHM), a vertex ranks the hyperedges containing it, a matching contains at most one hyperedge at each vertex, and an unchosen hyperedge is dominated by a vertex if that vertex prefers its matched hyperedge to it; otherwise the hyperedge blocks the matching~\cite{AF03}. In stable hypergraph $b$-matching (\SHbM), a vertex $v$ may lie in $b(v)$ chosen hyperedges, and a hyperedge is dominated if some vertex is saturated and prefers every chosen hyperedge to it~\cite{BF16}. Stable matchings need not exist, and deciding existence is NP-complete already for three-partite hypergraphs with hyperedges of size three~\cite{NH91}.

Scarf's lemma~\cite{Scarf67} guarantees a fractional stable solution~\cite{AF03,BF16}, but computing one is \PPAD-complete~\cite{KPRST13}, even under strong restrictions~\cite{IK18,Csa22}. Integrality gives a stronger existence guarantee. The matching polytope of a normal hypergraph is integral, so normal hypergraphs always admit stable matchings~\cite{AH98}, and unimodular hypergraphs, whose incidence matrices are totally unimodular, admit stable $b$-matchings for all integer capacities~\cite{BCS25}. The complexity of finding these integral solutions has remained a central question. Bir\'o, Cs\'aji and Schlotter~\cite{BCS25} gave a polynomial algorithm for \SHbM{} on laminar hypergraphs and for \SHM{} on subtree hypergraphs, and an XP algorithm for weighted \SHbM{} on subpath hypergraphs parameterized by the largest hyperedge. Faenza, He and Sethuraman~\cite{FHS26} proved polynomial convergence of a suitable implementation of Scarf's algorithm for bipartite stable matching. Chandrasekaran et al.~\cite{CFHS25} established polynomial convergence on arborescence hypergraphs, and Chandrasekaran and Kalathur~\cite{CK26} recently gave a polynomial algorithm for \SHM{} on all network hypergraphs. They leave open the unimodular and the normal cases, and the case of capacities on network hypergraphs.

\paragraph{Kernels in perfect graphs.}
A \emph{kernel} of a digraph is an independent set of vertices such that every other vertex has an arc into it. Kernels need not exist, and deciding their existence is NP-complete~\cite{Chv73}. A digraph is \emph{clique-acyclic} if every clique of its underlying graph induces an acyclic subdigraph. Confirming a conjecture of Berge and Duchet, Boros and Gurvich~\cite{BG96} proved that every clique-acyclic orientation of a perfect graph has a kernel (more generally, every clique-acyclic superorientation, where an edge may be oriented in both directions). Aharoni and Holzman~\cite{AH98} gave a short proof via Scarf's lemma. Stable matchings in a normal hypergraph are exactly the kernels of the conflict digraph on its hyperedges, whose underlying graph is the (perfect) line graph~\cite{AH98,KP10}. Computing such a kernel efficiently is a well-known open problem~\cite{Egres,PIM20}. Polynomial algorithms are known for several subclasses, such as chordal graphs and claw-free perfect graphs~\cite{PIM20}, line graphs of bipartite graphs~\cite{GS62}, and the conflict graphs of network hypergraphs~\cite{CK26}. 

\medskip
The University Dual Admission problem (\UDA) was introduced in~\cite{BCS25}, motivated by Hungarian university dual education~\cite{FFB19}. A student applies to programs, each program belongs to a university, and both universities and programs have capacities and rank students. A student must be accepted by both the program and its university. Bir\'o et al.~\cite{BCS25} showed that \UDA{} is a special case of \SHbM{} on network hypergraphs, so a stable assignment always exists. They also gave a linear-time algorithm when all capacities are one and left the general search complexity open.

\subsection{Our contribution}

We show that these integral solutions are \PPAD-hard to compute, even in very restricted settings.

\paragraph{Unit capacities: unimodular and normal hypergraphs, and kernels.}
Finding a stable matching is \PPAD-complete on unimodular hypergraphs, even if the hypergraph is linear, every hyperedge has at most three vertices, and every vertex lies in at most three hyperedges (Theorem~\ref{thm:unimodular}). Unimodular hypergraphs are balanced and normal, so the problem is \PPAD-complete on these classes as well. In the language of kernels, finding a kernel in a clique-acyclic orientation of a perfect graph is \PPAD-hard, even if the graph is the line graph of a unimodular hypergraph and has no clique of size four, and \PPAD-complete when the maximal cliques are given, in particular for bounded clique size (Corollary~\ref{cor:kernel}). This settles a long-standing open problem: unless $\PPAD\subseteq\FP$, there is no polynomial-time algorithm for kernels in perfect graphs, a question open since the existence theorems of Boros and Gurvich~\cite{BG96} and Aharoni and Holzman~\cite{AH98} and listed among the EGRES open problems~\cite{Egres}. It also settles the unimodular and normal cases of stable hypergraph matching, left open in~\cite{BCS25,CK26}. Together with~\cite{CK26}, it draws a sharp boundary inside the totally unimodular world: stable matchings are polynomial-time computable on network hypergraphs, but \PPAD-complete on unimodular hypergraphs.

\paragraph{Capacities: network hypergraphs and University Dual Admission.}
Finding a stable assignment in \UDA{} is \PPAD-complete, even when programs have quota one and at most two applicants, students apply to at most two programs at different universities, and university capacities are at most nine (Theorem~\ref{thm:main}). Consequently, \SHbM{} is \PPAD-complete on network hypergraphs, already with hyperedges of size three and capacities at most nine (Corollary~\ref{cor:shbm}). Thus capacities make even network hypergraphs hard, whereas their unit-capacity version is polynomial~\cite{CK26}, and \UDA{} with unit capacities is solvable in linear time~\cite{BCS25}.

\paragraph{Implications.}
Unless $\PPAD\subseteq\FP$, guaranteed existence of an integral stable solution does not imply a polynomial-time algorithm, even when the guarantee comes from total unimodularity. Our results concern an arbitrary stable solution, without any additional optimization objective. Earlier \PPAD-hardness results for hypergraph stability~\cite{KPRST13,IK18,Csa22}, including the strong fractional kernel problem~\cite{KPRST13}, concern fractional solutions. Our results transfer this difficulty to integral solutions of natural combinatorial problems on explicitly given graphs and hypergraphs, in classes whose matching polytope is integral.

Besides \UDA, the capacitated result applies to a subclass of the hospitals/residents problem with couples (\HRC). Interpret each student as a couple, universities as hospitals for the first member, and programs as hospitals for the second member. Every hospital on the program side is associated with one fixed hospital on the university side. In this subclass of dual markets, stable matchings always exist, but finding one is \PPAD-complete (Corollary~\ref{cor:hrc}). General dual markets do not have this existence guarantee~\cite{CMMT26}.

An important extension concerns near-feasible stable matchings. Nguyen and Vohra~\cite{NV18} proved that changing each hospital's capacity by at most two always suffices for a stable matching, while increasing total capacity by at most four. We prove that finding such capacities and a stable matching is \PPAD-hard (Corollary~\ref{cor:near-two}). More generally, hardness holds for every fixed additive allowance, even in the fixed-association subclass and with at most two acceptable pairs per couple (Theorem~\ref{thm:near}). Thus the hardness concerns the final near-feasible outcome, independently of the fractional solution used in their construction.

\paragraph{Techniques.}
Both reductions start from \PC~\cite{DFHM24} and represent every node of the circuit by a few students. Each student has a first-choice \emph{home} hyperedge in the gadget computing the node and a second-choice \emph{input} hyperedge in the gadget reading it. This wire mechanism makes gadgets interact only through whether these students are at home, which gives a clean local analysis. The reductions are short and modular, and we believe they provide a useful template for proving hardness of other structured problems with guaranteed integral solutions, such as further classes of hypergraphs, digraph kernels, stable flows and matching markets with couples. Section~\ref{sec:unimodular} isolates three reusable tools: dual-rail encoding with monotone gadgets, symmetry breaking through two coupled stable marriage markets, and a parity criterion that certifies total unimodularity of composed instances. The last one reduces checking unimodularity of an arbitrarily large instance to a finite check on each gadget.

\subsection{Proof overview}

The source problem \PC{} assigns values in $\{0,1,\bot\}$ to the nodes of a circuit. Its \NOR{} gates behave as Boolean NOR gates on pure values, while a \PUR{} gate copies a pure input to both outputs and must produce at least one pure output even from $\bot$. These two gate types suffice for \PPAD-completeness~\cite{DFHM24}. Additional \PUR{} gates ensure that each node is read at most once.

\emph{Capacities (Section~\ref{sec:uda}).} We represent a node by three students and decode zero when all three are at home, one when all three are away, and $\bot$ otherwise. Stability forces every student who is away from home to occupy her input program, so the number of arrivals to a gate is the number of displaced students of its inputs. Two small gadgets turn this number into the correct output by counting arguments, using that program and university priorities disagree.

\emph{Unit capacities (Section~\ref{sec:unimodular}).} Without capacities no counting is possible. We represent a node $z$ by two students, $z^1$ carrying the value and $z^2$ its negation (dual rail). A \NOR{} gate then only needs one \AND{} and one \OR{} of rails, both realized by tiny gadgets. The \PUR{} gate must break ties between the two codes $(0,0)$ and $(1,1)$ of $\bot$. The key gadget consists of two copies of the classical $2\times2$ stable marriage market with two stable matchings, coupled by a single hyperedge between two men; it never turns the input code $(0,0)$ into the output code $(0,0)$. Attaching negations to its four ports handles $(1,1)$. Finally, we show that every chordless cycle of the bipartite incidence graph of the composed hypergraph has length divisible by four. Since all degrees are at most three, Camion's characterization~\cite{Cam65} then shows that the incidence matrix is totally unimodular.

\paragraph{Organization.} Section~\ref{sec:prelim} contains preliminaries. Section~\ref{sec:uda} proves the result for \UDA{} and network $b$-matchings, together with the \HRC{} corollary. Section~\ref{sec:unimodular} builds on its wire mechanism and proves the result for unimodular hypergraphs, normal hypergraphs and kernels. Section~\ref{sec:near} treats near-feasible stable matchings with couples, and Section~\ref{sec:conclusion} concludes.

\section{Preliminaries}\label{sec:prelim}

\subsection{Stable hypergraph matching}

A hypergraph $\mathcal H=(V,E)$ has a finite vertex set $V$ and a finite family $E$ of nonempty subsets of $V$, called hyperedges. We write $E(v)=\{e\in E:v\in e\}$ for the \emph{star} of $v$ and call $|E(v)|$ its degree. The hypergraph is \emph{linear} if any two hyperedges share at most one vertex. Each vertex $v$ has a strict preference order $\succ_v$ on $E(v)$, with $e\succ_v f$ meaning that $v$ prefers $e$ to $f$. A capacity function $b:V\to\mathbb Z_{>0}$ specifies how many hyperedges a vertex may receive. A \emph{$b$-matching} is a set $M\subseteq E$ such that $|M(v)|\le b(v)$ for every $v$, where $M(v)=M\cap E(v)$. A vertex is \emph{full}, or \emph{saturated}, if equality holds.

A $b$-matching $M$ \emph{dominates} an unchosen hyperedge $e$ at $v\in e$ if $v$ is full and $f\succ_v e$ for every $f\in M(v)$. An unchosen hyperedge \emph{blocks} $M$ if it is dominated at none of its vertices. A $b$-matching is \emph{stable} if it has no blocking hyperedge. The search problem \SHbM{} asks for a stable $b$-matching. For $b\equiv1$, a $b$-matching is a \emph{matching}, and the corresponding problem is \SHM. For a matching $M$ we write $M(v)$ also for the hyperedge of $M$ containing $v$, if any; $v$ is \emph{covered} by $M$ if it exists.

The \emph{incidence matrix} of $\mathcal H$ is $A\in\{0,1\}^{V\times E}$ with $A_{ve}=1$ exactly when $v\in e$. A matrix is \emph{totally unimodular} (TU) if every square submatrix has determinant in $\{0,1,-1\}$. A hypergraph is \emph{unimodular} if its incidence matrix is TU. Its $b$-matching polytope
\[
 \mathcal P(\mathcal H,b)=\{x\in\mathbb R^E:Ax\le b,\ 0\le x\le1\}
\]
then has only integral vertices for every integer $b$~\cite{Schrijver}. A \emph{fractional $b$-matching} is a point $x\in\mathcal P(\mathcal H,b)$. It is \emph{stable} if, for every $e$ with $x_e<1$, some $v\in e$ satisfies $\sum_{f\in E(v)}x_f=b(v)$ and $x_f=0$ for every $f\in E(v)$ with $e\succ_v f$. For integral points this is precisely the stability condition above. Scarf's lemma guarantees such a fractional solution~\cite{AF03,BF16}.

A hypergraph is \emph{normal} if its unit-capacity matching polytope $\mathcal P(\mathcal H,\mathbf 1)$ is integral, equivalently if the edge chromatic number of every partial hypergraph equals its maximum degree~\cite{Lovasz72}.\footnote{A partial hypergraph is obtained by deleting hyperedges. The edge chromatic number is the fewest colors needed to color hyperedges so that intersecting hyperedges have different colors.} A \emph{cycle} of length $k\ge2$ is a sequence $v_1e_1v_2e_2\dots v_ke_kv_1$ of distinct vertices and distinct hyperedges with $v_i,v_{i+1}\in e_i$ (indices modulo $k$). A hypergraph is \emph{balanced} if in every cycle of odd length some $e_i$ contains at least three of $v_1,\dots,v_k$~\cite{Berge89}. Every unimodular hypergraph is balanced, and every balanced hypergraph is normal~\cite{Berge89}.

For an oriented tree $T$ and an ordered pair $(s,t)$ of its vertices, the vector of the unique tree path from $s$ to $t$ has entry $1$ on arcs traversed in their direction, $-1$ on arcs traversed against their direction, and $0$ elsewhere. A \emph{network matrix} has such path vectors as its columns and the arcs of $T$ as its rows. Network matrices are TU~\cite{Schrijver}. A \emph{network hypergraph} has a network matrix as its incidence matrix. Since this incidence matrix is nonnegative, each hyperedge corresponds to the arc set of a directed path in $T$. In particular, network hypergraphs are unimodular.

\paragraph{Bipartite incidence graph.}
The \emph{bipartite incidence graph} $B(\mathcal H)$ has vertex set $V\cup E$ and an edge $ve$ whenever $v\in e$; it is the bipartite graph of the matrix $A$. A \emph{hole} is an induced cycle and a \emph{chordless path} is an induced path. The length of a path or cycle is its number of edges. Holes of $B(\mathcal H)$ have even length. A cycle $v_1e_1\dots v_ke_kv_1$ of $\mathcal H$ is a cycle of length $2k$ in $B(\mathcal H)$, and no $e_i$ contains three of $v_1,\dots,v_k$ exactly when this cycle is a hole. Hence $\mathcal H$ is balanced if and only if every hole of $B(\mathcal H)$ has length divisible by four. We use the following characterization of total unimodularity.

\begin{theorem}[Camion~\cite{Cam65}]\label{thm:camion}
A matrix with entries in $\{0,1,-1\}$ is totally unimodular if and only if the sum of the entries of every square submatrix with even row and column sums is divisible by four.
\end{theorem}

\subsection{University Dual Admission}

In \UDA~\cite{BCS25}, there are finite disjoint sets of students $S$, universities $U$ and programs $P$. Each program $p$ belongs to exactly one university $u(p)$. Universities have positive integer capacities $c(u)$ and programs have positive integer quotas $q(p)$. Each student $s$ has a strict preference order $\succ_s$ over her acceptable programs and prefers every such program to being unassigned, denoted by $\emptyset$. Each university $u$ has a strict \emph{master order} $\succ_u$ over its applicants, and each program $p$ has a strict order $\succ_p$ over its applicants. Program and university orders may differ.

An assignment $\mu$ maps each student to an acceptable program or to $\emptyset$. Write $\mu(p)=\{s:\mu(s)=p\}$ and $\mu(u)=\bigcup_{p:u(p)=u}\mu(p)$. The assignment is \emph{feasible} if $|\mu(u)|\le c(u)$ and $|\mu(p)|\le q(p)$ for every university and program. A university or program is full if its constraint is tight.

\begin{definition}\label{def:uda-stability}
A feasible assignment is \emph{stable} if no acceptable triple $(s,u,p)$ with $u=u(p)$ satisfies all three conditions:
\begin{enumerate}[label=(\roman*),nosep]
\item $p\succ_s\mu(s)$;
\item $u$ is not full, or $s\succ_u s'$ for some $s'\in\mu(u)$, or $s\in\mu(u)$;
\item $p$ is not full, or $s\succ_p s''$ for some $s''\in\mu(p)$.
\end{enumerate}
Such a triple is a \emph{blocking triple}. The search problem \UDA{} asks for a stable assignment.
\end{definition}

The associated hypergraph has vertex set $S\cup U\cup P$ and a hyperedge $\{s,u(p),p\}$ for every acceptable student--program pair. Students have capacity one, universities capacity $c(u)$ and programs capacity $q(p)$. Student and program orders induce their hyperedge orders. A university first compares students by its master order and, for hyperedges of the same student, uses that student's order of programs. Stable assignments correspond bijectively to stable $b$-matchings in this hypergraph~\cite{BCS25}.

For completeness, its network representation has a vertex $r$, a vertex $v_u$ for each university, and vertices $v_p,v_s$ for programs and students. The tree arcs are $(v_s,r)$, $(r,v_u)$ and $(v_{u(p)},v_p)$, indexed by the corresponding agents. The path from $v_s$ to $v_p$ consists of the arcs indexed by $s,u(p),p$. Thus the hypergraph is a network hypergraph~\cite{BCS25} and every \UDA{} instance admits a stable assignment.

\begin{observation}\label{obs:reject}
If $\mu$ is stable and $p\succ_s\mu(s)$, where $p$ is a program of university $u$, then
\begin{enumerate}[label=(\Alph*),nosep]
\item\label{it:P} $p$ is full and prefers every student in $\mu(p)$ to $s$; or
\item\label{it:U} $u$ is full, $s\notin\mu(u)$, and $u$ prefers every student in $\mu(u)$ to $s$.
\end{enumerate}
Case~\ref{it:P} is impossible if $s$ is ranked first by $p$, and case~\ref{it:U} is impossible if $u$ is not full.
\end{observation}

\subsection{Total search and PPAD}

A polynomial search problem has solutions of polynomial encoding length that can be verified in polynomial time. It is \emph{total} if every input has a solution. The class \PPAD{} consists of the total search problems reducible to \textsc{End-of-Line}~\cite{Pap94}. This problem is given a directed graph, described by predecessor and successor circuits, with at most one incoming and one outgoing arc at each vertex, together with one known source. It asks for another source or a sink. A reduction constructs a target instance in polynomial time and converts any solution of that instance to a solution of the source instance in polynomial time. A problem is \PPAD-hard if every problem in \PPAD{} reduces to it, and \PPAD-complete if it is also in \PPAD. We write \FP{} for problems solvable in polynomial time.

\begin{proposition}\label{prop:membership}
Finding a stable $b$-matching in a unimodular hypergraph is in \PPAD. So is finding a stable matching in a normal hypergraph. Consequently, \UDA{} is in \PPAD.
\end{proposition}
\begin{proof}
Finding a fractional stable $b$-matching reduces to computing a dominating vertex of $\mathcal P(\mathcal H,b)$ in Scarf's lemma, with row rankings induced by the vertex preference orders, and this is in \PPAD~\cite{KPRST13}. If $\mathcal P(\mathcal H,b)$ is integral, which holds for unimodular hypergraphs and integer $b$ and for normal hypergraphs and $b\equiv1$, then the dominating vertex is integral and is a stable $b$-matching.
\end{proof}
As usual, the hypergraph class is a promise on the input; unimodularity can moreover be recognized in polynomial time by Seymour's decomposition~\cite{Sey80,Schrijver}.

\subsection{Pure-Circuit}

An instance of \PC~\cite{DFHM24} consists of a finite set of nodes $V$ and two types of gates. A \NOR{} gate $(\NOR,u,v,w)$ has inputs $u,v$ and output $w$. A \PUR{} gate $(\PUR,u,v,w)$ has input $u$ and outputs $v,w$. Within each gate $u,v,w$ are distinct, and every node is the output of exactly one gate. A node may be read by several gates or by none. Directed cycles are allowed.

A solution $x:V\to\{0,1,\bot\}$ satisfies the following conditions:
\begin{enumerate}[label=(\roman*),nosep]
\item For a \NOR{} gate, $x[u]=x[v]=0$ implies $x[w]=1$, and $1\in\{x[u],x[v]\}$ implies $x[w]=0$. Otherwise its output is unrestricted.
\item For a \PUR{} gate, at least one of $x[v],x[w]$ belongs to $\{0,1\}$, and $x[u]\in\{0,1\}$ implies $x[v]=x[w]=x[u]$.
\end{enumerate}
The values zero and one are called \emph{pure}. Deligkas et al.~\cite{DFHM24} proved that finding a solution is \PPAD-complete.

\begin{lemma}\label{lem:fanout}
We may assume that the \PC{} instance is fan-out free, that is, every node is an input of at most one gate.
\end{lemma}
\begin{proof}
If a node $u$ is read by $k\ge2$ gates, put $r_0=u$ and add
\[
 (\PUR,r_{i-1},u_i,r_i)\quad(1\le i\le k-2),\qquad
 (\PUR,r_{k-2},u_{k-1},u_k),
\]
where all $u_i$ and $r_i$ for $i\ge1$ are newly created helper nodes. Let the $i$-th reader of $u$ read input $u_i$. The number of added gates is linear in the number of original input occurrences. Every node is now read at most once, and a pure value of $u$ is copied to every $u_i$.

Restrict any solution to the original nodes. For an original \NOR{} gate, every pure input triggering a condition is copied to the input of the modified gate, so its original output satisfies that condition. For an original \PUR{} gate, its two outputs are unchanged and at least one is pure, while a pure original input is copied to its modified input. Thus the restriction solves the original instance.
\end{proof}

\section{Stable \texorpdfstring{$b$}{b}-matchings in network hypergraphs}\label{sec:uda}

We first prove hardness for \UDA{} (Theorem~\ref{thm:main}); the results for network hypergraphs and for couples then follow from the network representation of Section~\ref{sec:prelim}. Fix a fan-out free \PC{} instance. We construct a \UDA{} instance as follows.

\paragraph{Node students.} For each node $z$, we add three students $z^1,z^2,z^3$. Each student $z^j$ has a dedicated \emph{home program} $\home(z^j)$, which she ranks first, in the gadget of the gate that outputs $z$. If $z$ is read by a gate, she also has a dedicated \emph{input program} $\inp(z^j)$, which she ranks second, in the gadget of that gate. That is, $z^j$ has preference $\home(z^j)\succ \inp(z^j)$. A student is \emph{at home} if she is assigned to her home program and \emph{away} otherwise.

\paragraph{The \NOR{} gadget.} Let $g=(\NOR,u,v,w)$. The input students are $u^j,v^j$ ($j=1,2,3$). We create a university $U_g$
with capacity $9$ and programs $\inp(u^j),\inp(v^j),d_1,\dots,d_7$, as well as a university $N_g$ that has capacity $3$ and
programs $f_1,f_2,f_3,\home(w^1),\home(w^2),\home(w^3)$. We create dummy students $D_k$ ($k\le7$) that apply only to $d_k$, and
reservoir students $T_k$ ($k\le3$) with preference $d_k\succ f_k$. For $k\le3$, program $d_k$ ranks $D_k\succ T_k$. The master
orders are
\begin{align*}
U_g&:\ u^1\succ u^2\succ u^3\succ v^1\succ v^2\succ v^3\succ T_1\succ T_2\succ T_3\succ D_7\succ\dots\succ D_1,\\
N_g&:\ T_1\succ T_2\succ T_3\succ w^3\succ w^2\succ w^1.
\end{align*}

\paragraph{The \PUR{} gadget.} Let $g=(\PUR,u,v,w)$. The input students are $u^1,u^2,u^3$. A single university $U_g$
is created with capacity $7$ and programs $\inp(u^j),\home(v^j),\home(w^j)$. For $x\in\{v,w\}$ and $j\le2$, reservoir students
$R_{x,j}$ are added too, who apply only to $\home(x^j)$. The priority order of $\home(x^j)$ is $x^j\succ R_{x,j}$. The master order is
\begin{align*}
U_g:\quad&
u^1\succ u^2\succ u^3\succ R_{v,1}\succ R_{v,2}\succ R_{w,1}\succ R_{w,2}\\
&{}\succ w^3\succ w^2\succ w^1\succ v^3\succ v^2\succ v^1.
\end{align*}

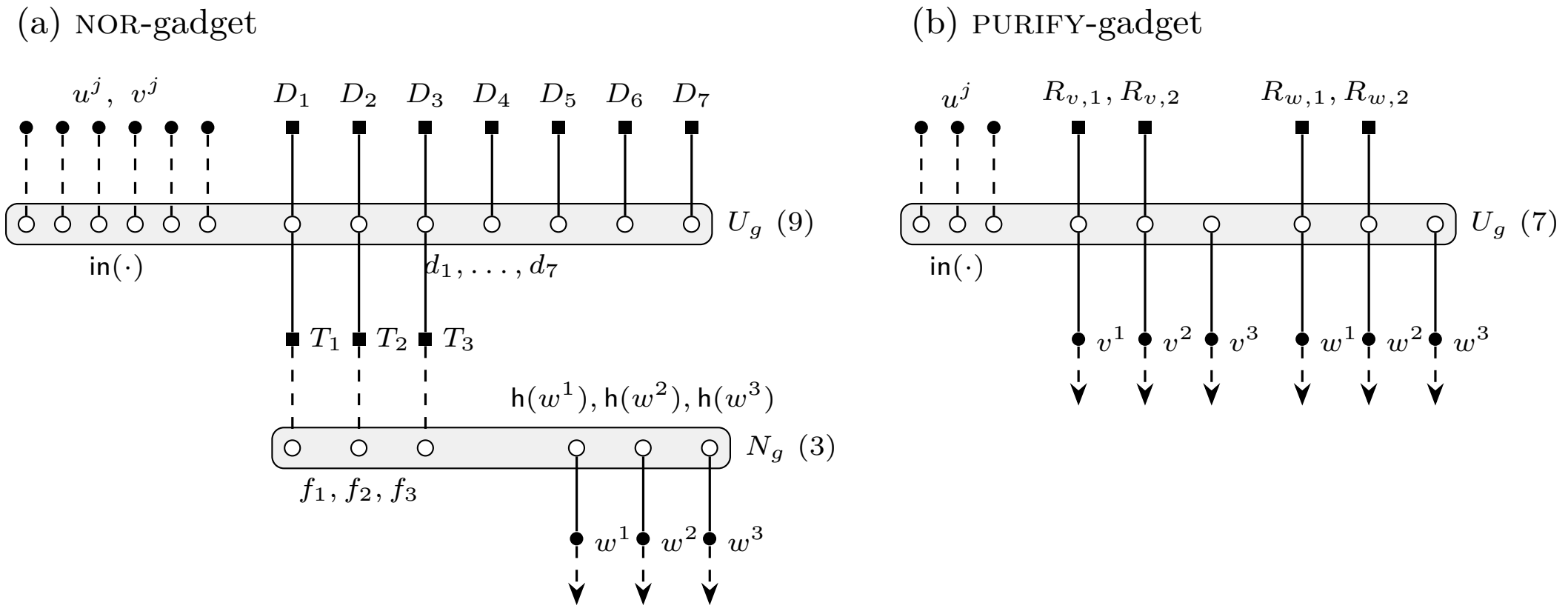
\begin{figure}[t]
\centering
\begin{tikzpicture}[scale=1.08,
  stu/.style={circle,fill=black,inner sep=1.2pt},
  res/.style={rectangle,fill=black,inner sep=1.6pt},
  prog/.style={circle,draw,fill=white,inner sep=1.4pt},
  first/.style={semithick},
  second/.style={semithick,dashed},
  univ/.style={draw,rounded corners=3pt,fill=gray!12,inner sep=3pt},
  lab/.style={font=\scriptsize},
  >=Stealth]
\begin{scope}
  \foreach \i in {1,...,6}{
    \node[prog] (in\i) at (0.3*\i-0.3,2.6) {};
    \node[stu] (s\i) at (0.3*\i-0.3,3.4) {};
    \draw[second] (s\i) -- (in\i);
  }
  \node[lab,above] at (0.75,3.45) {$u^j,\ v^j$};
  \foreach \k in {1,...,7}{
    \node[prog] (d\k) at (1.65+0.55*\k,2.6) {};
    \node[res,label={[lab]above:$D_{\k}$}] (D\k) at (1.65+0.55*\k,3.4) {};
    \draw[first] (D\k) -- (d\k);
  }
  \begin{scope}[on background layer]
    \node[univ,fit=(in1)(d7),label={[font=\scriptsize]right:$U_g$ (9)}] {};
  \end{scope}
  \node[lab,below] at (3.85,2.47) {$d_1,\dots,d_7$};
  \node[lab,below] at (0.75,2.47) {$\inp(\cdot)$};
  \foreach \k in {1,2,3}{
    \node[res,label={[lab,xshift=-1pt]right:$T_{\k}$}] (T\k) at (1.65+0.55*\k,1.65) {};
    \node[prog] (f\k) at (1.65+0.55*\k,0.75) {};
    \draw[first] (T\k) -- (d\k);
    \draw[second] (T\k) -- (f\k);
  }
  \foreach \j in {1,2,3}{
    \node[prog] (h\j) at (4.0+0.55*\j,0.75) {};
    \node[stu,label={[lab,xshift=-1pt]right:$w^{\j}$}] (w\j) at (4.0+0.55*\j,0) {};
    \draw[first] (w\j) -- (h\j);
    \draw[second,->] (w\j) -- ++(0,-0.55);
  }
  \begin{scope}[on background layer]
    \node[univ,fit=(f1)(h3),label={[font=\scriptsize]right:$N_g$ (3)}] {};
  \end{scope}
  \node[lab,below] at (2.75,0.62) {$f_1,f_2,f_3$};
  \node[lab,above] at (5.1,0.92) {$\home(w^1),\home(w^2),\home(w^3)$};
  \node[font=\small,anchor=west] at (-0.2,4.25) {(a) \NOR-gadget};
\end{scope}
\begin{scope}[xshift=7.4cm]
  \foreach \i in {1,2,3}{
    \node[prog] (pi\i) at (0.3*\i-0.3,2.6) {};
    \node[stu] (pu\i) at (0.3*\i-0.3,3.4) {};
    \draw[second] (pu\i) -- (pi\i);
  }
  \node[lab,above] at (0.3,3.45) {$u^j$};
  \node[lab,below] at (0.3,2.47) {$\inp(\cdot)$};
  \foreach \x/\o in {v/0.75,w/2.6}{
    \foreach \j in {1,2,3}{
      \node[prog] (h\x\j) at (\o+0.55*\j,2.6) {};
      \node[stu,label={[lab,xshift=-1pt]right:$\x^{\j}$}] (o\x\j) at (\o+0.55*\j,1.65) {};
      \draw[first] (o\x\j) -- (h\x\j);
      \draw[second,->] (o\x\j) -- ++(0,-0.55);
    }
    \foreach \j in {1,2}{
      \node[res] (R\x\j) at (\o+0.55*\j,3.4) {};
      \draw[first] (R\x\j) -- (h\x\j);
    }
    \node[lab,above] at (\o+0.825,3.45) {$R_{\x,1},R_{\x,2}$};
  }
  \begin{scope}[on background layer]
    \node[univ,fit=(pi1)(hw3),label={[font=\scriptsize]right:$U_g$ (7)}] {};
  \end{scope}
  \node[font=\small,anchor=west] at (-0.2,4.25) {(b) \PUR-gadget};
\end{scope}
\end{tikzpicture}
\caption{The \UDA{} gadgets. Shaded boxes are universities (capacity in parentheses), white circles are programs of
quota~$1$, dots are node students and squares are gadget students. Solid lines are first choices of students, dashed lines
second choices; in (b) the programs joined to $v^j,w^j$ are $\home(v^j),\home(w^j)$. Input students have their
homes elsewhere, and the dashed arrows of output students lead to the
gate reading the node.}
\label{fig:gadgets}
\end{figure}

All programs have quota $1$. Gadget students, programs and universities are distinct across gates; the node students are shared between the gate that produces a node and the gate that reads it. Every program whose order has not been specified has only its named applicant, whom it ranks first. Figure~\ref{fig:gadgets} shows both gadgets. The construction is polynomial. Every student applies to at most two
programs, at different universities, and every program has at most two applicants.

\medskip
 For a node $z$, let
$d(z)$ be the number of students of $z$ who are away from home. We decode $x[z]=0$ if $d(z)=0$, $x[z]=1$ if
$d(z)=3$, and $x[z]=\bot$ otherwise. For a gate $g$, let $a_g$ be the number of its input students collectively at their input
programs.

\paragraph{Correctness.} Every constructed instance admits a stable assignment by its network representation. Fix such an assignment $\mu$.

\begin{lemma}\label{lem:arrive}
Every input student of a gate $g$ is at home or at its input program. Hence $a_g=d(u)+d(v)$ for a \NOR-gate and
$a_g=d(u)$ for a \PUR-gate.
\end{lemma}

\begin{proof}
Suppose an input student $s$ is unassigned. She is the only applicant of $\inp(s)$, so by
Observation~\ref{obs:reject} the university $U_g$ is full of students ranked above $s$. Only other input
students rank above $s$, and there are at most $5<9$ (\NOR) or $2<7$ (\PUR) of them.
\end{proof}

\begin{lemma}\label{lem:nor}
Let $g=(\NOR,u,v,w)$ and $a=a_g$. If $a\le1$ then $d(w)=3$. If $a\ge3$ then $d(w)=0$.
\end{lemma}

\begin{proof}
All quotas are $1$, so $|\mu(U_g)|\le a+7$.

\emph{Case $a\le1$.} Then $U_g$ is not full. Each $D_k$ is first at $d_k$, so $D_k$ is at $d_k$ by
Observation~\ref{obs:reject}. Each $T_k$ is then not at $d_k$. It is the only applicant of $f_k$, and at most
two students rank above it at $N_g$, so $T_k$ is at $f_k$. Thus $N_g$ is full with $T_1,T_2,T_3$, and $d(w)=3$.

\emph{Case $a\ge3$.} Let $k$ be minimal with $T_k$ not at $d_k$. At most $6+2=8<9$ students rank above $T_k$ at
$U_g$, so by Observation~\ref{obs:reject} $D_k$ is at $d_k$. Each $D_m$ with $m>k$ is first at $d_m$, and $U_g$
contains $D_k\prec_{U_g}D_m$, so by stability $D_m$ is at $d_m$ as well. Then
\[
|\mu(U_g)|\ge a+(k-1)+(8-k)=a+7\ge10>9,
\]
a contradiction. Hence every $T_k$ is at $U_g$. Then $N_g$ contains only $w$-students and is not full if some
$w^j$ is away. Since $w^j$ is the only applicant of $\home(w^j)$, every $w^j$ is at home and $d(w)=0$.
\end{proof}

\begin{lemma}\label{lem:pur}
Let $g=(\PUR,u,v,w)$ and $a=a_g$. If $a=0$ then $d(v)=d(w)=0$; if $a=1$ then $d(w)=0$; if $a=2$ then
$d(v)=3$; and if $a=3$ then $d(v)=d(w)=3$.
\end{lemma}

\begin{proof}
List the output students bottom-up in the master order of $U_g$ as
\[
y_1,\dots,y_6=v^1,v^2,v^3,w^1,w^2,w^3 .
\]
Each $y_i$ is first at its home program, so by Observation~\ref{obs:reject}:

\smallskip
\noindent(F1) if $y_i$ is away, then $U_g$ is full and ranks all of $\mu(U_g)$ above $y_i$.
\smallskip

\noindent Hence if $y_i$ is at home, so is every $y_{i'}$ with $i'>i$, because $y_i\in\mu(U_g)$ ranks below
$y_{i'}$. The away students are therefore
$y_1,\dots,y_t$ for some $t$, and $d(v)=\min(t,3)$, $d(w)=\max(t-3,0)$.

Now let $R_{x,j}$ be a reservoir student whose corresponding $x^j$ is away. Then nobody at $\home(x^j)$ is preferred to $R_{x,j}$,
and at most $6<7$ students rank above $R_{x,j}$ at $U_g$. So by Observation~\ref{obs:reject}, $R_{x,j}$ is at
$\home(x^j)$. Therefore
\[
|\mu(U_g)|=a+(6-t)+r_t,
\]
where $r_t$ counts the reservoirs at the homes of $y_1,\dots,y_t$, so that $(r_0,\dots,r_6)=(0,1,2,2,3,4,4)$.
Feasibility requires $|\mu(U_g)|\le7$, and (F1) requires equality when $t\ge1$. Hence
\begin{itemize}[nosep]
\item $t=0$ needs $a+6\le7$;
\item $t\in\{1,2\}$ needs $a+6=7$;
\item $t\in\{3,4,5\}$ needs $a+5=7$;
\item $t=6$ needs $a+4=7$.
\end{itemize}
So $a=0$ forces $t=0$, $a=1$ forces $t\le2$, $a=2$ forces $3\le t\le5$, and $a=3$ forces $t=6$.
\end{proof}

\begin{theorem}\label{thm:main}
Finding a stable assignment in \UDA{} is \PPAD-complete. This holds even if all quotas are $1$, programs have at
most two applicants, students apply to at most two programs at different universities, and university capacities are at most $9$.
\end{theorem}

\begin{proof}
Membership follows from Proposition~\ref{prop:membership}. For hardness, decode a stable assignment as above, and use
Lemmas~\ref{lem:arrive}--\ref{lem:pur}. For a \NOR-gate, $x[u]=x[v]=0$ gives $a_g=0$ and hence
$x[w]=1$, while $1\in\{x[u],x[v]\}$ gives $a_g\ge3$ and hence $x[w]=0$. For a \PUR-gate, a pure input gives
$a_g\in\{0,3\}$, and both outputs copy it. In every case $a_g\le1$ makes $x[w]=0$ and $a_g\ge2$ makes $x[v]=1$,
so an output is pure. Hence $x$ is a solution, and \PC{} reduces to \UDA.
\end{proof}

\subsection{Consequences: network hypergraphs and couples}

\begin{corollary}\label{cor:shbm}
\SHbM{} is \PPAD-complete on network hypergraphs and on unimodular hypergraphs, already with hyperedges of size $3$
and capacities at most $9$.
\end{corollary}

\begin{proof}
The network representation in Section~\ref{sec:prelim} transforms every instance constructed for Theorem~\ref{thm:main} into a three-uniform \SHbM{} instance with the same capacity bound. Network hypergraphs are unimodular, and membership follows from Proposition~\ref{prop:membership}.
\end{proof}

The hospitals/residents problem with couples (\HRC) also has an interpretation of the reduction. In general, deciding whether an \HRC{} instance admits a stable matching is NP-complete~\cite{Ronn90,MM10}; see~\cite{BK13} for a survey. Scarf-based methods have been used to search for stable matchings~\cite{BFI16} and to obtain stable outcomes after modifying hospital capacities~\cite{NV18}.

Here we consider instances consisting entirely of couples. Each couple ranks acceptable ordered pairs of hospitals and prefers every such pair to being unassigned. Each hospital has a positive integer capacity and a strict order on its applicants. A feasible matching assigns both members of a couple to an acceptable pair or leaves both unassigned, and respects hospital capacities. We restrict attention to \emph{dual markets}, with disjoint hospital sets $H_1,H_2$, where first members apply only to $H_1$ and second members only to $H_2$. A couple assigned to $(h_0,h'_0)$ blocks with a preferred pair $(h,h')$ if each proposed hospital either already holds the corresponding member, has a free seat, or prefers that member to a current resident. The couple has no current hospitals if it is unassigned.

We impose the additional restriction that a function $\phi:H_2\to H_1$ is fixed and every acceptable pair is of the form $(\phi(h'),h')$. Thus each hospital on the second side is associated with a single hospital on the first side. This is a subclass of dual markets; unrestricted dual markets may have no stable matching~\cite{CMMT26}.

\begin{corollary}\label{cor:hrc}
Every \HRC{} instance with this fixed hospital association admits a stable matching. Finding one is \PPAD-complete, even when each couple ranks at most two pairs and all capacities are at most nine.
\end{corollary}
\begin{proof}
Map a \UDA{} student $s$ to a couple $(s_U,s_P)$, universities to $H_1$, and programs to $H_2$, with $\phi(p)=u(p)$. Hospital $u$ ranks $s_U$ by its master order, hospital $p$ ranks $s_P$ by $\succ_p$, and the couple ranks $(u(p),p)$ as $s$ ranks $p$. The same construction in reverse transforms any instance with the fixed association into a \UDA{} instance. Feasibility is preserved. In a preferred pair, the first member's hospital already holding her is exactly the condition $s\in\mu(u)$ in Definition~\ref{def:uda-stability}. The second member's current hospital cannot be the proposed program, since then the couple would already hold the proposed pair. The blocking conditions therefore coincide. Existence follows from the network representation, and completeness from Theorem~\ref{thm:main} and Proposition~\ref{prop:membership}.
\end{proof}

\FloatBarrier
\section{Stable matchings in unimodular hypergraphs}\label{sec:unimodular}

We now show that unit capacities do not make the problem easy once we move from network hypergraphs to unimodular ones. Following the terminology of Section 3, we refer to the vertices representing circuit values as students. In gadgets based on classical stable marriage markets, we additionally refer to certain internal vertices as men and women. These terms are used solely to aid intuition; formally, all agents are vertices of a hypergraph with unit capacities. As in Section~\ref{sec:uda}, we reduce from a fan-out free \PC{} instance, and every node is represented by students with a first-choice home hyperedge and a second-choice input hyperedge. Since there are no capacities to count with, we use two students per node instead of three, in a dual-rail encoding, and build all gates from small gadgets. Section~\ref{sec:wiring} sets up the wiring framework, Sections~\ref{sec:basic}--\ref{sec:break} describe the gadgets, Section~\ref{sec:reduction2} gives the reduction, and Section~\ref{sec:tu} proves that the constructed hypergraph is unimodular.

\subsection{Gadgets and wires}\label{sec:wiring}

A \emph{gadget} $X$ is a hypergraph with vertex preferences together with two disjoint sets of vertices, its \emph{input ports} and \emph{output ports}. Every input port $a$ lies in exactly one hyperedge of $X$, its \emph{input edge} $\inp(a)$, and every output port $y$ lies in exactly one hyperedge of $X$, its \emph{home edge} $\home(y)$. No hyperedge contains two ports.

Given gadgets with pairwise disjoint vertex sets, a \emph{wiring} identifies some output ports with input ports of \emph{other} gadgets, such that every port is identified with at most one other port. The identified vertex $y$ is a \emph{wire}; it lies in exactly the two hyperedges $\home(y)$ and $\inp(y)$, of the two different gadgets, and ranks $\home(y)\succ_y\inp(y)$. All other vertices keep their hyperedges and preferences. Ports that are not identified remain \emph{external} ports of the resulting hypergraph, which is therefore again a gadget.

\paragraph{Home and away.}
For a matching $M$ and a port $y$, we say that $y$ is \emph{at home} if $\home(y)\in M$ and \emph{away} otherwise, and we write $r(y)=0$ or $r(y)=1$, respectively. In the reduction, these values of the rail students encode the circuit values. An away input port $a$ may be covered by $\inp(a)$ or be uncovered; $r(a)=1$ only says that $\inp(a)$ is available to $a$.

\paragraph{Structure of the argument.}
The hypergraph $\mathcal H$ constructed in Section~\ref{sec:reduction2} is unimodular (Theorem~\ref{thm:unimodular}), so it has a stable matching by Proposition~\ref{prop:membership}. We never construct a stable matching, and we never have to check that a given matching is stable. Instead we fix an \emph{arbitrary} stable matching $M$ of $\mathcal H$ and derive necessary conditions on the values $r(\cdot)$, one gadget at a time. This is done in two separate steps.
\begin{enumerate}[label=(\arabic*),nosep]
\item \emph{Gadget relations.} Each gadget $X$ is analysed in isolation. For every input vector $\alpha$, we take an arbitrary stable matching of the small hypergraph $X[\alpha]$ defined below and show that its outputs satisfy a relation $R_X$ (Lemmas~\ref{lem:basic} and~\ref{lem:break0}). Here stability is a \emph{hypothesis}: the proofs only use that no hyperedge blocks.
\item \emph{Locality.} Lemma~\ref{lem:local} \emph{proves} that the restriction of a stable matching of the whole hypergraph to a gadget $X$ is a stable matching of $X[\alpha]$, where $\alpha$ consists of the values $r(a)$ of the input ports of $X$. This is the only place where stability of a matching is proved.
\end{enumerate}
Together, (1) and (2) show that the values $r(\cdot)$ in $M$ satisfy the relations of all gadgets simultaneously. Lemma~\ref{lem:gates} then reads off the circuit values.

\paragraph{The isolated gadget $X[\alpha]$.}
Let $X$ be a gadget with input ports $I$ and output ports $O$, and let $\alpha\in\{0,1\}^I$. Let $X[\alpha]$ be the hypergraph obtained from $X$ by deleting $\inp(a)$ for every $a\in I$ with $\alpha_a=0$, with the preferences of $X$ restricted to the remaining hyperedges. The vector $\alpha$ models everything that $X$ sees of the rest of the instance.
\begin{itemize}[nosep]
\item An input port $a$ with $\alpha_a=0$ is at home in the gadget producing it and prefers $\home(a)$ to $\inp(a)$. So $\inp(a)$ is neither in the matching nor blocking, and it can be deleted.
\item An input port $a$ with $\alpha_a=1$ is not at home. Inside $X$ its only hyperedge is $\inp(a)$, which it would accept.
\item An output port $y$ lies only in $\home(y)$ inside $X$. Its input edge in the reading gadget is its second choice, so for $X$ it behaves like a vertex whose only hyperedge is $\home(y)$.
\end{itemize}
For a matching of $X[\alpha]$ and an output port $y$, let $r_y=0$ if $y$ is covered, necessarily by $\home(y)$, and $r_y=1$ otherwise. This agrees with $r(y)$ defined above.

\paragraph{Computing a relation.}
We say that $X$ \emph{computes} a relation $R\subseteq\{0,1\}^I\times\{0,1\}^O$ if for every $\alpha\in\{0,1\}^I$ and every stable matching of $X[\alpha]$, the pair $(\alpha,r)$ belongs to $R$. When $R$ is the graph of a function, we say that $X$ computes this function. This is a statement about \emph{all} stable matchings of $X[\alpha]$. It does not claim that $X[\alpha]$ has one, and it may allow several outputs for the same $\alpha$, as for \BRK0 with $\alpha=(0,0)$.

\begin{lemma}[Locality]\label{lem:local}
Let $Y$ be a wiring of gadgets, $\alpha$ a vector on the external input ports of $Y$, and $M$ a stable matching of $Y[\alpha]$. For each gadget $X$ of the wiring, define $\alpha^X$ on the input ports of $X$ by $\alpha^X_a=\alpha_a$ if $a$ is an external input port of $Y$, and $\alpha^X_a=r(a)$ if $a$ is a wire. Then:
\begin{enumerate}[label=(\roman*),nosep]
\item $M_X=M\cap E(X)$ is a stable matching of $X[\alpha^X]$;
\item an output port $y$ of $X$ is covered by $M_X$ if and only if $r(y)=0$.
\end{enumerate}
Consequently, if every gadget $X$ computes a relation $R_X$, then the values $r(\cdot)$ satisfy all relations $R_X$ simultaneously.
\end{lemma}

We use Lemma~\ref{lem:local} in two ways. With $Y=\mathcal H$, which has no external ports, it transfers the gadget relations to an arbitrary stable matching of $\mathcal H$. Furthermore, with $Y$ being a small composite gadget, such as \BRK1 or the \NOR{} and \PUR{} gadgets, it shows that $Y$ computes the relation obtained by composing the relations of its parts (Lemmas~\ref{lem:break1} and~\ref{lem:gates}).

\begin{proof}
The hypothesis is that $M$ is stable in $Y[\alpha]$; we prove that $M_X$ is stable in $X[\alpha^X]$.

\emph{$M_X$ is a matching of $X[\alpha^X]$.} We check that $M_X$ contains no deleted hyperedge. For an external input port $a$ with $\alpha_a=0$, the edge $\inp(a)$ is not in $Y[\alpha]$, so it is not in $M$. For a wire $a$ with $r(a)=0$, we have $\home(a)\in M$, so $\inp(a)\notin M$.

\emph{No hyperedge of $X[\alpha^X]$ blocks $M_X$.} Let $e$ be a hyperedge of $X[\alpha^X]$ with $e\notin M_X$. Then $e$ is a hyperedge of $Y[\alpha]$ not in $M$. Since $M$ is stable in $Y[\alpha]$, $e$ does not block $M$, so some $v\in e$ is covered by a hyperedge $M(v)\succ_v e$. We show that $M(v)\in M_X$, so that $v$ also dominates $e$ with respect to $M_X$ in $X[\alpha^X]$.
\begin{itemize}[nosep]
\item $v$ is not an output port $y$ of $X$, because then $e=\home(y)$, the first choice of $y$.
\item $v$ is not an input port $a$ of $X$. If $a$ is external, then $e=\inp(a)$ is its only hyperedge in $Y[\alpha]$. If $a$ is a wire, then $M(a)\succ_a\inp(a)=e$ forces $M(a)=\home(a)$, so $\alpha^X_a=r(a)=0$ and $e$ was deleted from $X[\alpha^X]$.
\item Hence $v$ is an internal vertex of $X$. All hyperedges at $v$ belong to $X$, so $M(v)\in M_X$.
\end{itemize}

\emph{Outputs.} An output port $y$ of $X$ lies only in $\home(y)$ inside $X$, so it is covered by $M_X$ if and only if $\home(y)\in M$, that is, $r(y)=0$.

The final claim follows by applying the definition of computing $R_X$ to the stable matching $M_X$ of $X[\alpha^X]$, whose input vector is $\alpha^X$ and whose output values are $r(\cdot)$ by (ii).
\end{proof}

\paragraph{How gadget relations are proved.}
In the proofs of Lemmas~\ref{lem:basic} and~\ref{lem:break0}, we fix $\alpha$ and an arbitrary stable matching $M$ of $X[\alpha]$, and we use stability only through the following consequence: \emph{if a hyperedge $e$ is not in $M$, then some vertex of $e$ is covered by a hyperedge it prefers to $e$.} In particular, if every vertex of $e$ ranks $e$ first among its hyperedges in $X[\alpha]$, then $e\in M$. These arguments determine what $M$ must look like; they never require checking that some matching is stable.

\subsection{Monotone gadgets}\label{sec:basic}

Table~\ref{tab:basic} lists four basic gadgets. A vertex that lies in a single hyperedge has a trivial preference order and is omitted. The gadget \NOT{} is used only inside the gadget \BRK{1} of Section~\ref{sec:break}.

\begin{table}[!ht]
\centering
\small
\begin{tabular}{@{}llll@{}}
\toprule
Gadget & Hyperedges & Preferences & Computes \\
\midrule
\COPY & $\inp(a)=\{a,s_1,s_2\}$,\ $\home(y_j)=\{s_j,y_j\}$ & $s_j:\ \inp(a)\succ\home(y_j)$ & $y_1=y_2=a$\\[2pt]
\OR & $\inp(a_i)=\{a_i,s_i\}$,\ $\home(y)=\{s_1,s_2,y\}$ & $s_i:\ \inp(a_i)\succ\home(y)$ & $y=a_1\vee a_2$\\[2pt]
\AND & $\inp(a_i)=\{a_i,c_i\}$,\ $b_i=\{c_i,q_i\}$, & $c_i:\ \inp(a_i)\succ b_i$,\ \ $q_i:\ b_i\succ p$, & $y=a_1\wedge a_2$\\
 & $p=\{q_1,q_2,s\}$,\ $\home(y)=\{s,y\}$ & $s:\ p\succ\home(y)$ & \\[2pt]
\NOT & $\inp(a)=\{a,z\}$,\ $g=\{z,t\}$,\ $\home(y)=\{t,y\}$ & $z:\ \inp(a)\succ g$,\ \ $t:\ g\succ\home(y)$ & $y=1-a$\\
\bottomrule
\end{tabular}
\caption{The basic gadgets, with $i,j\in\{1,2\}$. Input ports are $a,a_1,a_2$ and output ports are $y,y_1,y_2$. The last column states the computed function, writing a port for its value $r(\cdot)$.}
\label{tab:basic}
\end{table}

\begin{lemma}\label{lem:basic}
The gadgets \COPY, \OR, \AND{} and \NOT{} compute the functions in Table~\ref{tab:basic}.
\end{lemma}

\begin{proof}
Fix $\alpha$ and a stable matching $M$ of $X[\alpha]$. Recall that an input port $a$ with $\alpha_a=1$ lies only in $\inp(a)$.

\emph{\COPY.} If $\alpha_a=1$, every vertex of $\inp(a)$ ranks it first, so $\inp(a)\in M$, and $y_1,y_2$ are uncovered. If $\alpha_a=0$, then $\home(y_j)$ is the only hyperedge of $s_j$ and of $y_j$, so $\home(y_j)\in M$ for $j=1,2$.

\emph{\OR.} If $\alpha_{a_i}=1$ for some $i$, then $\inp(a_i)\in M$ as before, so $s_i$ is not available for $\home(y)$ and $y$ is uncovered. If $\alpha=(0,0)$, then $\home(y)$ is the only hyperedge of all its vertices, so $\home(y)\in M$.

\emph{\AND.} If $\alpha_{a_i}=1$, then $\inp(a_i)\in M$, so $b_i\notin M$ and $q_i$ can only be covered by $p$. If $\alpha_{a_i}=0$, then $b_i$ is the only hyperedge of $c_i$ and the first choice of $q_i$, so $b_i\in M$ and $p\notin M$. Hence if $\alpha=(1,1)$, every vertex of $p$ is uncovered or covered by $p$, and $s$ ranks $p$ first, so $p\in M$ and $y$ is uncovered. Otherwise $p\notin M$, so $\home(y)$ is the only remaining hyperedge of $s$ and $y$, and $\home(y)\in M$.

\emph{\NOT.} If $\alpha_a=1$, then $\inp(a)\in M$, so $g\notin M$, and $\home(y)$ is the only remaining hyperedge of $t$ and $y$, so $\home(y)\in M$. If $\alpha_a=0$, then $g$ is the only hyperedge of $z$ and the first choice of $t$, so $g\in M$ and $y$ is uncovered.
\end{proof}

\subsection{Breaking ties with two marriage markets}\label{sec:break}

The only non-monotone behaviour we need comes from a gadget built from two copies of the classical $2\times2$ marriage market with two stable matchings. We write $xy$ for a hyperedge $\{x,y\}$.

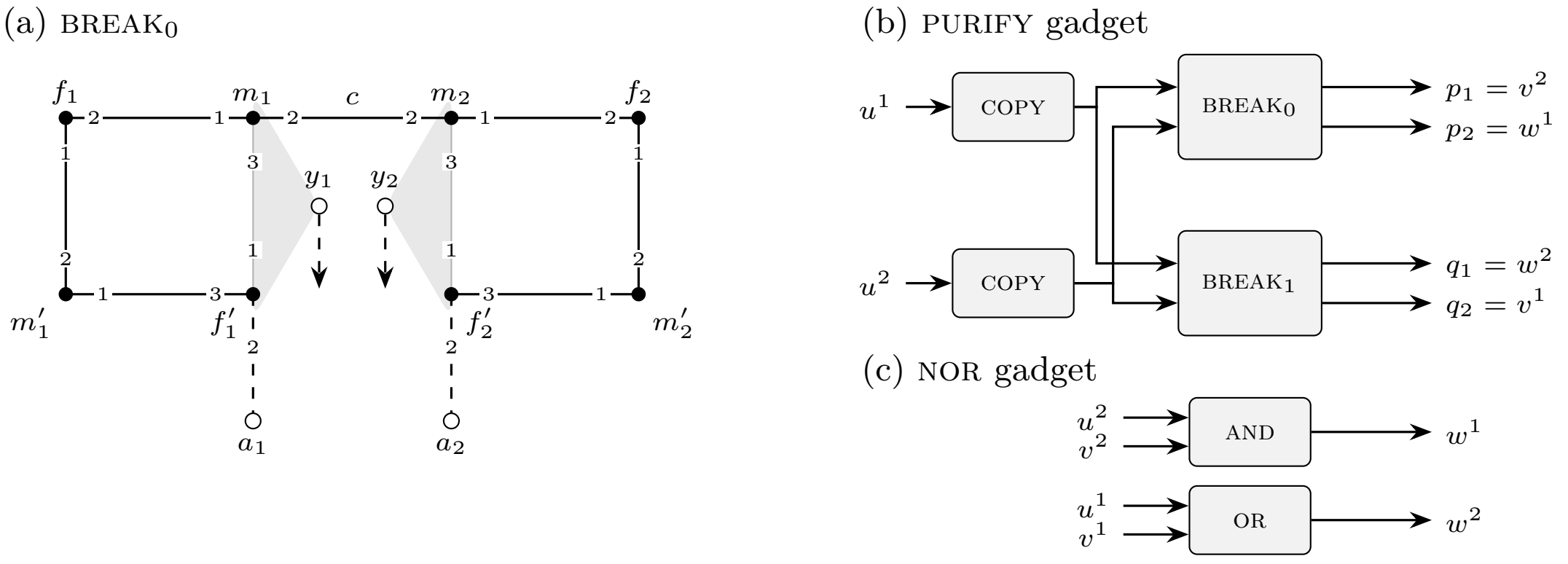
\begin{figure}[t]
\centering
\begin{tikzpicture}[
  ag/.style={circle,fill=black,inner sep=1.3pt},
  port/.style={circle,draw,fill=white,inner sep=1.4pt},
  lab/.style={font=\scriptsize},
  rk/.style={font=\tiny,fill=white,inner sep=0.6pt},
  first/.style={semithick},
  inp/.style={semithick,dashed},
  >=Stealth]
\begin{scope}

  \foreach \s/\x in {1/-1,2/1}{
    \coordinate (M\s) at (\x*0.9,1.4);
    \coordinate (F\s) at (\x*2.6,1.4);
    \coordinate (Mp\s) at (\x*2.6,-0.2);
    \coordinate (Fp\s) at (\x*0.9,-0.2);
    \coordinate (Y\s) at (\x*0.3,0.6);
    \coordinate (A\s) at (\x*0.9,-1.35);
  }
  \foreach \s in {1,2}{
    \fill[gray!18,rounded corners=4pt] ($(M\s)+(0,0.22)$) -- ($(Fp\s)+(0,-0.22)$) -- ($(Y\s)$) -- cycle;
  }
  \foreach \s in {1,2}{
    \draw[first] (M\s) -- (F\s);
    \draw[first] (F\s) -- (Mp\s);
    \draw[first] (Mp\s) -- (Fp\s);
    \draw[gray!60] (M\s) -- (Fp\s);
    \draw[inp] (Fp\s) -- (A\s);
  }
  \draw[first] (M1) -- (M2);

  \foreach \s in {1,2}{
    \node[ag] at (M\s) {}; \node[ag] at (F\s) {}; \node[ag] at (Mp\s) {}; \node[ag] at (Fp\s) {};
    \node[port] (yy\s) at (Y\s) {}; \node[port] (aa\s) at (A\s) {};
  }
  \node[lab,above] at (M1) {$m_1$}; \node[lab,above] at (M2) {$m_2$};
  \node[lab,above] at (F1) {$f_1$}; \node[lab,above] at (F2) {$f_2$};
  \node[lab,below left] at (Mp1) {$m'_1$}; \node[lab,below right] at (Mp2) {$m'_2$};
  \node[lab,below left] at (Fp1) {$f'_1$}; \node[lab,below right] at (Fp2) {$f'_2$};
  \node[lab,below=1pt] at (A1) {$a_1$}; \node[lab,below=1pt] at (A2) {$a_2$};
  \node[lab,above=1pt] at (Y1) {$y_1$}; \node[lab,above=1pt] at (Y2) {$y_2$};
  \node[lab,above] at (0,1.4) {$c$};
  \draw[inp,->] (yy1) -- ++(0,-0.75);
  \draw[inp,->] (yy2) -- ++(0,-0.75);

  \foreach \s/\d in {1/-1,2/1}{
    \node[rk] at ($(M\s)!0.18!(F\s)$) {1};
    \node[rk] at ($(F\s)!0.15!(M\s)$) {2};
    \node[rk] at ($(F\s)!0.2!(Mp\s)$) {1};
    \node[rk] at ($(Mp\s)!0.2!(F\s)$) {2};
    \node[rk] at ($(Mp\s)!0.2!(Fp\s)$) {1};
    \node[rk] at ($(Fp\s)!0.2!(Mp\s)$) {3};
    \node[rk] at ($(Fp\s)!0.25!(M\s)$) {1};
    \node[rk] at ($(Fp\s)!0.42!(A\s)$) {2};
    \node[rk] at ($(M\s)!0.25!(Fp\s)$) {3};
  }
  \node[rk] at ($(M1)!0.2!(M2)$) {2};
  \node[rk] at ($(M2)!0.2!(M1)$) {2};
  \node[font=\small,anchor=west] at (-3.3,2.25) {(a) \BRK0};
\end{scope}
\begin{scope}[xshift=7.6cm,yshift=0.2cm,
  box/.style={draw,rounded corners=2pt,fill=gray!10,minimum height=0.62cm,inner sep=2pt,font=\scriptsize},
  wire/.style={semithick,->}]
  \node[box,minimum width=1.1cm] (C1) at (-1.6,1.3) {\COPY};
  \node[box,minimum width=1.1cm] (C2) at (-1.6,-0.3) {\COPY};
  \node[box,minimum width=1.3cm,minimum height=0.95cm] (B0) at (0.55,1.3) {\BRK0};
  \node[box,minimum width=1.3cm,minimum height=0.95cm] (B1) at (0.55,-0.3) {\BRK1};
  \node[lab] (u1) at (-2.85,1.3) {$u^1$};
  \node[lab] (u2) at (-2.85,-0.3) {$u^2$};
  \draw[wire] (u1) -- (C1); \draw[wire] (u2) -- (C2);
  \draw[wire] (C1.east) -- ++(0.2,0) |- ($(B0.west)+(0,0.18)$);
  \draw[wire] (C1.east) -- ++(0.2,0) |- ($(B1.west)+(0,0.18)$);
  \draw[wire] (C2.east) -- ++(0.35,0) |- ($(B0.west)+(0,-0.18)$);
  \draw[wire] (C2.east) -- ++(0.35,0) |- ($(B1.west)+(0,-0.18)$);
  \node[lab,anchor=west] (o1) at (2.2,1.48) {$p_1=v^2$};
  \node[lab,anchor=west] (o2) at (2.2,1.12) {$p_2=w^1$};
  \node[lab,anchor=west] (o3) at (2.2,-0.12) {$q_1=w^2$};
  \node[lab,anchor=west] (o4) at (2.2,-0.48) {$q_2=v^1$};
  \draw[wire] ($(B0.east)+(0,0.18)$) -- (o1.west);
  \draw[wire] ($(B0.east)+(0,-0.18)$) -- (o2.west);
  \draw[wire] ($(B1.east)+(0,0.18)$) -- (o3.west);
  \draw[wire] ($(B1.east)+(0,-0.18)$) -- (o4.west);
  \node[font=\small,anchor=west] at (-3.1,2.05) {(b) \PUR{} gadget};

  \node[box,minimum width=1.1cm] (AN) at (0.55,-1.65) {\AND};
  \node[box,minimum width=1.1cm] (ORG) at (0.55,-2.45) {\OR};
  \node[lab,anchor=east] (i1) at (-0.6,-1.52) {$u^2$};
  \node[lab,anchor=east] (i2) at (-0.6,-1.78) {$v^2$};
  \node[lab,anchor=east] (i3) at (-0.6,-2.32) {$u^1$};
  \node[lab,anchor=east] (i4) at (-0.6,-2.58) {$v^1$};
  \draw[wire] (i1.east) -- ($(AN.west)+(0,0.13)$);
  \draw[wire] (i2.east) -- ($(AN.west)+(0,-0.13)$);
  \draw[wire] (i3.east) -- ($(ORG.west)+(0,0.13)$);
  \draw[wire] (i4.east) -- ($(ORG.west)+(0,-0.13)$);
  \node[lab,anchor=west] (n1) at (2.2,-1.65) {$w^1$};
  \node[lab,anchor=west] (n2) at (2.2,-2.45) {$w^2$};
  \draw[wire] (AN) -- (n1.west); \draw[wire] (ORG) -- (n2.west);
  \node[font=\small,anchor=west] at (-3.1,-1.1) {(c) \NOR{} gadget};
\end{scope}
\end{tikzpicture}
\caption{(a) The gadget \BRK0. Dots are gadget vertices and white circles are ports. Black segments are hyperedges of size two, and the shaded triangles are the home edges $\home(y_i)=\{m_i,f'_i,y_i\}$. Small numbers give the rank of a hyperedge in the preference order of the adjacent vertex. Dashed segments are input edges $\inp(a_i)=a_if'_i$, and dashed arrows lead to the gadgets reading $y_1,y_2$. (b) The \PUR{} gadget for $(\PUR,u,v,w)$; \BRK1 is \BRK0 with a \NOT{} gadget at each port. Ports of \BRK0 and \BRK1 are drawn in the order $a_1,a_2$ (left) and $y_1,y_2$ (right). (c) The \NOR{} gadget for $(\NOR,u,v,w)$.}
\label{fig:break}
\end{figure}

\paragraph{The gadget \BRK0.} It has input ports $a_1,a_2$ and output ports $y_1,y_2$. For $i=1,2$, side $i$ consists of two men $m_i,m_i'$ and two women $f_i,f_i'$ with the four hyperedges
\[
 m_if_i,\qquad \home(y_i)=\{m_i,f_i',y_i\},\qquad m_i'f_i',\qquad m_i'f_i,
\]
and the input edge $\inp(a_i)=a_if_i'$. A single \emph{coupling} hyperedge $c=m_1m_2$ joins the two sides. The preferences are
\begin{align*}
 m_i&:\ m_if_i\succ c\succ\home(y_i), & f_i'&:\ \home(y_i)\succ\inp(a_i)\succ m_i'f_i',\\
 m_i'&:\ m_i'f_i'\succ m_i'f_i, & f_i&:\ m_i'f_i\succ m_if_i.
\end{align*}
Ignoring $c$, $y_i$ and $a_i$, for each side $i$ we get a marriage market in which both men rank different women first and both women rank different men first. It has exactly two stable matchings: the men-optimal $A_i=\{m_if_i,\ m_i'f_i'\}$ and the women-optimal $B_i=\{\home(y_i),\ m_i'f_i\}$. In $A_i$ the output $y_i$ is away and $m_i$ has his first choice; in $B_i$ the output $y_i$ is at home and $m_i$ would rather take the coupling edge. An arriving input $a_i$ destroys $A_i$, since $f_i'$ prefers $\inp(a_i)$ to $m_i'f_i'$. Figure~\ref{fig:break}(a) shows the gadget.

\begin{lemma}\label{lem:break0}
\BRK0 computes the relation
\[
 (y_1,y_2)=(a_2,a_1)\ \text{ if } (a_1,a_2)\ne(0,0),\qquad (y_1,y_2)\ne(0,0)\ \text{ if } (a_1,a_2)=(0,0).
\]
\end{lemma}

\begin{proof}
Fix $\alpha$ and a stable matching $M$ of $\BRK0[\alpha]$.

\emph{Claim 1: if $c\notin M$, then $M$ contains $A_i$ or $B_i$ for each side $i$, and it contains $A_i$ only if $\alpha_{a_i}=0$.}

Suppose first $m_i'f_i'\in M$. Then $\home(y_i)\notin M$, as it meets $f_i'$, and $m_i'f_i\notin M$, as it meets $m_i'$. If $m_if_i\notin M$, then $m_i$ and $f_i$ are both uncovered, since also $c\notin M$, and $m_if_i$ blocks. Thus $A_i\subseteq M$. If $\alpha_{a_i}=1$, then $\inp(a_i)$ blocks, since $a_i$ is uncovered and $f_i'$ prefers $\inp(a_i)$ to $m_i'f_i'$; hence $\alpha_{a_i}=0$.

Suppose now $m_i'f_i'\notin M$. Then $m_i'f_i\in M$, since otherwise $m_i'$ is uncovered and $f_i$ ranks $m_i'f_i$ first, so $m_i'f_i$ blocks. Now $m_if_i\notin M$ and $c\notin M$, so $m_i$ is uncovered unless $\home(y_i)\in M$. The vertex $y_i$ lies only in $\home(y_i)$, and $f_i'$ ranks $\home(y_i)$ first. Hence $\home(y_i)\in M$, as otherwise it blocks, and $B_i\subseteq M$.

\emph{Claim 2: if $c\in M$, then $\alpha=(1,1)$ and $y_1,y_2$ are uncovered.}
The edges $\home(y_i)$ meet $c$ at $m_i$, so $y_1,y_2$ are uncovered. Suppose $\alpha_{a_i}=0$. If $m_i'f_i'\in M$, then $f_i$ is uncovered, as $m_if_i$ meets $c$ and $m_i'f_i$ meets $m_i'f_i'$; so $m_if_i$ blocks because $m_i$ prefers it to $c$. If $m_i'f_i\in M$, then $f_i'$ is uncovered, since her other hyperedges are $\home(y_i)$, which meets $c$, and the deleted $\inp(a_i)$; so $m_i'f_i'$ blocks. If $m_i'$ is uncovered, then $m_i'f_i$ blocks. Each case contradicts stability.

\emph{Conclusion.} If $c\notin M$, then $c$ does not block, so some $m_i$ prefers his hyperedge in $M$ to $c$, that is, not both sides use $B_i$. For $\alpha=(0,0)$, Claim~2 gives $c\notin M$, so some side uses $A_i$ and its $y_i$ is uncovered; hence $(y_1,y_2)\ne(0,0)$. For $\alpha=(1,0)$, Claim~2 gives $c\notin M$, Claim~1 gives $B_1\subseteq M$, and then $A_2\subseteq M$; so $(y_1,y_2)=(0,1)$. The case $\alpha=(0,1)$ is symmetric. For $\alpha=(1,1)$, Claim~1 would put $B_1\cup B_2$ in $M$ if $c\notin M$, which is impossible as $c$ would block; so $c\in M$ and $(y_1,y_2)=(1,1)$ by Claim~2.
\end{proof}

All three outcomes for $\alpha=(0,0)$ occur: $A_1\cup A_2$, $A_1\cup B_2$ and $B_1\cup A_2$ are stable. The relation of Lemma~\ref{lem:break0} is therefore not monotone, and it is exactly what is needed to purify the code $(0,0)$.

\paragraph{The gadget \BRK1.} Take a copy of \BRK0 and four \NOT{} gadgets. For $i=1,2$, wire the output of the $i$-th \NOT{} gadget to the input port $a_i$ of \BRK0, and the output port $y_i$ of \BRK0 to the input of the $(2+i)$-th \NOT{} gadget. The input ports of \BRK1 are the inputs $a_1',a_2'$ of the first two \NOT{} gadgets, and its output ports are the outputs $y_1',y_2'$ of the last two.

\begin{lemma}\label{lem:break1}
\BRK1 computes the relation
\[
 (y'_1,y'_2)=(a'_2,a'_1)\ \text{ if } (a'_1,a'_2)\ne(1,1),\qquad (y'_1,y'_2)\ne(1,1)\ \text{ if } (a'_1,a'_2)=(1,1).
\]
\end{lemma}

\begin{proof}
By Lemmas~\ref{lem:local} and~\ref{lem:basic}, $a_i=1-a'_i$ and $y'_i=1-y_i$, where $(a,y)$ satisfies Lemma~\ref{lem:break0}. Complementing that relation gives the claim.
\end{proof}

\subsection{The reduction}\label{sec:reduction2}

Fix a fan-out free \PC{} instance. For each node $z$, we create two \emph{rail students} $z^1,z^2$. We decode
\[
 x[z]=\begin{cases}
 1,& (r(z^1),r(z^2))=(1,0),\\
 0,& (r(z^1),r(z^2))=(0,1),\\
 \bot,& \text{otherwise}.
 \end{cases}
\]
Thus $z^1$ carries the value of $z$ and $z^2$ its negation, and both codes $(0,0)$ and $(1,1)$ mean $\bot$. For each gate we create a gadget whose input ports are the rail students of its inputs and whose output ports are the rail students of its outputs; Figure~\ref{fig:break}(b,c) shows them.

\begin{itemize}[nosep,leftmargin=*]
\item For $g=(\NOR,u,v,w)$: an \AND{} gadget with input ports $u^2,v^2$ and output port $w^1$, and an \OR{} gadget with input ports $u^1,v^1$ and output port $w^2$.
\item For $g=(\PUR,u,v,w)$: two \COPY{} gadgets, one with input port $u^1$ and one with input port $u^2$; a \BRK0 gadget whose input $a_j$ is wired to the first output of the \COPY{} gadget of $u^j$; and a \BRK1 gadget whose input $a'_j$ is wired to the second output of the \COPY{} gadget of $u^j$. Writing $(p_1,p_2)$ and $(q_1,q_2)$ for the output ports of \BRK0 and \BRK1, the rail students of the outputs are
\[
 v^1=q_2,\qquad v^2=p_1,\qquad w^1=p_2,\qquad w^2=q_1.
\]
\end{itemize}
Finally, wire every rail student to the gadget of the gate reading its node, if any. Let $\mathcal H$ be the resulting hypergraph. A \NOR{} gadget has $9$ hyperedges and a \PUR{} gadget has $40$, so the construction is polynomial. Every wire joins two different gadgets: inside a \PUR{} gadget this is clear, and between gates it follows from the distinctness of $u,v,w$ within each gate. Every rail student has exactly one home edge, and at most one input edge because the instance is fan-out free.

\begin{lemma}\label{lem:gates}
Let $M$ be a stable matching of $\mathcal H$ and $x$ its decoding.
\begin{enumerate}[label=(\alph*),nosep]
\item For a gate $(\NOR,u,v,w)$, $x[u]=x[v]=0$ implies $x[w]=1$, and $1\in\{x[u],x[v]\}$ implies $x[w]=0$.
\item For a gate $(\PUR,u,v,w)$, $x[u]\in\{0,1\}$ implies $x[v]=x[w]=x[u]$. Moreover, $x[v]=0$ or $x[w]=1$.
\end{enumerate}
\end{lemma}

\begin{proof}
By Lemma~\ref{lem:local}, the values $r(\cdot)$ satisfy the relations of all gadgets.

(a) By Lemma~\ref{lem:basic}, $r(w^1)=r(u^2)\wedge r(v^2)$ and $r(w^2)=r(u^1)\vee r(v^1)$. If $x[u]=x[v]=0$, then $r(u^1)=r(v^1)=0$ and $r(u^2)=r(v^2)=1$, so $(r(w^1),r(w^2))=(1,0)$ and $x[w]=1$. If $x[u]=1$, then $r(u^1)=1$ and $r(u^2)=0$, so $(r(w^1),r(w^2))=(0,1)$ and $x[w]=0$; the case $x[v]=1$ is symmetric.

(b) Both inputs of \BRK0 and of \BRK1 receive $(r(u^1),r(u^2))$ by Lemma~\ref{lem:basic}. Lemmas~\ref{lem:break0} and~\ref{lem:break1} give the following cases, where $p=(r(p_1),r(p_2))$ and $q=(r(q_1),r(q_2))$, and $x[v]$, $x[w]$ are decoded from $(r(q_2),r(p_1))$ and $(r(p_2),r(q_1))$.
\begin{center}
\small
\begin{tabular}{@{}cccccc@{}}
\toprule
$(r(u^1),r(u^2))$ & $x[u]$ & $p$ & $q$ & $(r(v^1),r(v^2))$ & $(r(w^1),r(w^2))$\\
\midrule
$(0,1)$ & $0$ & $(1,0)$ & $(1,0)$ & $(0,1)$ & $(0,1)$\\
$(1,0)$ & $1$ & $(0,1)$ & $(0,1)$ & $(1,0)$ & $(1,0)$\\
$(0,0)$ & $\bot$ & $\ne(0,0)$ & $(0,0)$ & $(0,r(p_1))$ & $(r(p_2),0)$\\
$(1,1)$ & $\bot$ & $(1,1)$ & $\ne(1,1)$ & $(r(q_2),1)$ & $(1,r(q_1))$\\
\bottomrule
\end{tabular}
\end{center}
A pure input is copied to both outputs, so $x[u]=0$ gives $x[v]=0$ and $x[u]=1$ gives $x[w]=1$. For input $(0,0)$, either $r(p_1)=1$ and $x[v]=0$, or $r(p_2)=1$ and $x[w]=1$. For input $(1,1)$, either $r(q_2)=0$ and $x[v]=0$, or $r(q_1)=0$ and $x[w]=1$.
\end{proof}

Lemma~\ref{lem:gates} shows that the decoding of any stable matching of $\mathcal H$ is a solution of the \PC{} instance. It remains to show that $\mathcal H$ is unimodular.

\subsection{Unimodularity}\label{sec:tu}

For hypergraphs of bounded degree, Camion's theorem reduces total unimodularity to a parity condition on holes.

\begin{lemma}\label{lem:camion}
Let $\mathcal H$ be a hypergraph in which every hyperedge has at most three vertices and every vertex lies in at most three hyperedges. If every hole of $B(\mathcal H)$ has length divisible by four, then $\mathcal H$ is unimodular.
\end{lemma}

\begin{proof}
Let $A'$ be a square submatrix of the incidence matrix with even row and column sums. Each row and column of $A$ has at most three ones, so each row and column of $A'$ has zero or two ones. Delete the zero rows and columns. The ones of the remaining matrix form a $2$-regular subgraph of $B(\mathcal H)$, which is the subgraph induced by its rows and columns. Hence it is a disjoint union of holes of $B(\mathcal H)$, and the sum of the entries of $A'$ is the sum of their lengths, which is divisible by four. Theorem~\ref{thm:camion} completes the proof.
\end{proof}

The next lemma shows that the parity of holes can be checked gadget by gadget. For a gadget $X$, let $B(X)$ be its bipartite incidence graph, with the ports among its vertices. Consider the following two conditions:
\begin{enumerate}[label=(H\arabic*),nosep]
\item\label{H1} every hole of $B(X)$ has length divisible by four;
\item\label{H2} every chordless path of $B(X)$ between two distinct ports has length divisible by four if exactly one of the two ports is an input port, and length congruent to $2$ modulo $4$ otherwise.
\end{enumerate}

\begin{lemma}\label{lem:parity}
Let $\mathcal H$ be a wiring of gadgets that satisfy \ref{H1} and \ref{H2}. Then every hole of $B(\mathcal H)$ has length divisible by four.
\end{lemma}

\begin{proof}
Every edge $ve$ of $B(\mathcal H)$ belongs to $B(X)$ for the unique gadget $X$ containing $e$. Let $C$ be a hole of $B(\mathcal H)$. If all its edges belong to one gadget $X$, then $C$ is a hole of $B(X)$ and~\ref{H1} applies. Otherwise, call a vertex of $C$ a \emph{switch} if its two edges on $C$ belong to different gadgets. A switch lies in two gadgets, so it is a wire $y$, and its two neighbours on $C$ are $\home(y)$ and $\inp(y)$. The switches cut $C$ into paths $P_1,\dots,P_k$, each contained in $B(X)$ for a single gadget $X$ and joining two ports of $X$. If there were a single switch $y$, then all edges of $C$ would lie in one gadget, since the gadget changes only at switches, contradicting that $y$ is a switch. Hence $k\ge2$, and each $P_j$ has two distinct ends. Each $P_j$ is chordless in $B(X)$, as $C$ is a hole.

Give a port the label $+1$ in a gadget where it is an input port and $-1$ in a gadget where it is an output port. Condition~\ref{H2} says that the length of $P_j$ is congruent modulo four to the sum of the labels of its two ends in its gadget: $0$ for an input and an output, $2$ for two inputs, and $-2\equiv2$ for two outputs. Every switch $y$ is an end of exactly two paths, one in the gadget where $y$ is an output port and one in the gadget where it is an input port. Therefore
\[
 |C|=\sum_{j=1}^k|P_j|\equiv\sum_{\text{switches } y}\bigl((-1)+(+1)\bigr)=0\pmod 4.\qedhere
\]
\end{proof}

\begin{lemma}\label{lem:gadget-parity}
The gadgets \COPY, \OR, \AND, \BRK0 and \BRK1 satisfy \ref{H1} and \ref{H2}.
\end{lemma}

\begin{proof}
For \COPY, \OR{} and \AND, the graph $B(X)$ is a tree, so~\ref{H1} holds and each pair of ports is joined by a unique path. The input--output paths have lengths $4$ (\COPY{} and \OR, e.g.\ $a,\inp(a),s_1,\home(y_1),y_1$) and $8$ (\AND: $a_i,\inp(a_i),c_i,b_i,q_i,p,s,\home(y),y$). The paths between the two outputs of \COPY{} and between the two inputs of \OR{} have length $6$, and between the two inputs of \AND{} length $10$.

For \BRK0, side $i$ forms the cycle
\[
 C_i:\ m_i,\ m_if_i,\ f_i,\ m_i'f_i,\ m_i',\ m_i'f_i',\ f_i',\ \home(y_i),\ m_i
\]
of length $8$. The rest of $B(\BRK0)$ consists of the path $m_1,c,m_2$ and the pendant paths $a_i,\inp(a_i),f_i'$ and $y_i,\home(y_i)$. Thus the holes are $C_1$ and $C_2$, and~\ref{H1} holds. A chordless path from $a_i$ starts with $a_i,\inp(a_i),f_i'$, and one from $y_i$ starts with $y_i,\home(y_i)$. Paths between the two sides pass through $m_1,c,m_2$.
 
The only remaining choice is how a path runs through $C_i$, and here we use that a chordless path is induced: any two of its vertices that are adjacent in $B(\BRK0)$ must be consecutive on the path. The vertex of $\home(y_i)$ is adjacent exactly to $m_i$, $f_i'$ and $y_i$. So if a chordless path contains $\home(y_i)$ together with $m_i$, or together with $f_i'$, then these two vertices are consecutive on it. This rules out every detour around $C_i$ that ends at $\home(y_i)$:
\begin{enumerate}[label=(\alph*),nosep]
\item A chordless path from $f_i'$ to $m_i$ is either $f_i',\home(y_i),m_i$, of length $2$, or the other way around $C_i$, namely $f_i',m_i'f_i',m_i',m_i'f_i,f_i,m_if_i,m_i$, of length $6$. The second one avoids $\home(y_i)$.
\item A chordless path from $f_i'$ to $y_i$ is $f_i',\home(y_i),y_i$. Going around $C_i$ to $m_i$ first and then to $\home(y_i)$ would contain $f_i'$ and $\home(y_i)$ without being consecutive.
\item Symmetrically, a chordless path from $m_i$ to $y_i$ is $m_i,\home(y_i),y_i$, since going around $C_i$ to $f_i'$ first would contain $m_i$ and $\home(y_i)$ without being consecutive.
\end{enumerate}
Combining the initial segments at $a_i$ and $y_i$, the passage $m_1,c,m_2$ between the sides, and (a)--(c), the chordless paths between ports have the following lengths:
$a_i$--$y_i$: $4$; $a_i$--$y_{3-i}$: $2+\{2,6\}+2+2\in\{8,12\}$; $a_1$--$a_2$: $2+\{2,6\}+2+\{2,6\}+2\in\{10,14,18\}$; $y_1$--$y_2$: $6$. This proves~\ref{H2}.

Finally, $B(\BRK1)$ is obtained from $B(\BRK0)$ by attaching at each port a pendant path of length $6$ through a \NOT{} gadget, for instance $a_i',\inp(a_i'),z,g,t,\home(a_i),a_i$. This creates no new hole and adds $12$ to the length of every chordless path between ports.
\end{proof}

A single \NOT{} gadget violates~\ref{H2}: its input--output path has length $6$. This is why negation in the circuit is realized by the choice of rails, and \NOT{} gadgets occur only inside \BRK1, where every path between two of its ports passes through exactly two of them.

\begin{theorem}\label{thm:unimodular}
Finding a stable matching in a unimodular hypergraph is \PPAD-complete. This holds even for linear hypergraphs in which every hyperedge has at most three vertices and every vertex lies in at most three hyperedges.
\end{theorem}

\begin{proof}
Membership follows from Proposition~\ref{prop:membership}. For hardness, construct $\mathcal H$ as in Section~\ref{sec:reduction2}. It is a wiring of \COPY, \OR, \AND, \BRK0 and \BRK1 gadgets, so Lemmas~\ref{lem:parity} and~\ref{lem:gadget-parity} show that every hole of $B(\mathcal H)$ has length divisible by four. Every hyperedge has at most three vertices. A rail student lies in at most two hyperedges, the vertices $m_i$ and $f_i'$ of a \BRK0 gadget in three, and every other vertex in at most two. So $\mathcal H$ is unimodular by Lemma~\ref{lem:camion}. Each gadget is linear by inspection of Table~\ref{tab:basic} and the definition of \BRK0, and hyperedges of different gadgets meet only in a wire, so $\mathcal H$ is linear. Any stable matching decodes in polynomial time to a solution of the \PC{} instance by Lemma~\ref{lem:gates}.
\end{proof}

\subsection{Normal hypergraphs and kernels in perfect graphs}\label{sec:kernels}

Unimodular hypergraphs are balanced and normal~\cite{Berge89}, and Proposition~\ref{prop:membership} applies to normal hypergraphs. Hence Theorem~\ref{thm:unimodular} gives the following.

\begin{corollary}\label{cor:normal}
Finding a stable matching is \PPAD-complete on balanced hypergraphs and on normal hypergraphs.
\end{corollary}

We now translate the result to kernels. Let $D$ be a digraph. A \emph{kernel} of $D$ is an independent set $K$ such that every vertex outside $K$ has an arc to some vertex of $K$. The digraph is an \emph{orientation} of a graph $G$ if every edge of $G$ is oriented in exactly one direction, and it is \emph{clique-acyclic} if every clique of $G$ induces an acyclic subdigraph. Boros and Gurvich~\cite{BG96} proved that every clique-acyclic orientation of a perfect graph has a kernel. In the proof of Aharoni and Holzman~\cite{AH98}, Scarf's lemma applied to the maximal-clique constraints yields a dominating vertex of the polytope $\{x\ge0:\ x(Q)\le1 \text{ for every maximal clique } Q\}$; this polytope is integral for perfect graphs, and the integral dominating vertex is a kernel. With the maximal cliques given as input, this places the search problem in \PPAD{} by~\cite{KPRST13}; as usual for such promise problems, we assume that the input graph is perfect and that the list of maximal cliques is complete.

\begin{corollary}\label{cor:kernel}
Finding a kernel in a clique-acyclic orientation of a perfect graph is \PPAD-hard, even if the graph is the line graph of a unimodular hypergraph and has no clique on four vertices. It is \PPAD-complete when the maximal cliques are part of the input, in particular for graphs of bounded clique number.
\end{corollary}

\begin{proof}
Let $\mathcal H$ be an instance of Theorem~\ref{thm:unimodular}, and let $D$ be the digraph with vertices corresponding to the hyperedges $e\in E$ with an arc $e\to f$ whenever $e\cap f=\{v\}$ and $f\succ_v e$. Since $\mathcal H$ is linear, $D$ is an orientation of the line graph $G=L(\mathcal H)$. A set $K\subseteq E$ is independent in $G$ if and only if it is a matching, and every $e\notin K$ has an arc into $K$ if and only if every $e\notin K$ is dominated. Hence the kernels of $D$ are exactly the stable matchings of $\mathcal H$.

\emph{Cliques.} Three pairwise intersecting hyperedges without a common vertex would meet in three distinct vertices $v_1,v_2,v_3$ by linearity and form a hole $e_1,v_3,e_2,v_1,e_3,v_2,e_1$ of length $6$ in $B(\mathcal H)$, contradicting Lemmas~\ref{lem:parity} and~\ref{lem:gadget-parity}. So every triangle of $G$ lies in a star $E(v)$, and by linearity so does every clique: if $e_1,e_2,e_3$ share $v$, then any further $e_4$ of the clique contains $e_1\cap e_2=\{v\}$. Consequently, cliques have at most three vertices, and $D$ restricted to a clique $Q\subseteq E(v)$ follows the order $\succ_v$, so it is acyclic.

\emph{Perfectness.} For $F\subseteq E$, the hypergraph $(V,F)$ has a TU incidence matrix, so linear programming duality and integrality give $\nu(F)=\tau(F)$, where $\nu(F)$ is the maximum number of disjoint hyperedges of $F$ and $\tau(F)$ the minimum number of vertices meeting all of them. In $G[F]$, $\nu(F)$ is the independence number, and since cliques lie in stars, $\tau(F)$ is the minimum number of cliques covering $F$. Hence the complement of $G$ is perfect, and so is $G$ by the perfect graph theorem~\cite{Lovasz72}.

The construction of $D$ is polynomial and kernels translate back to stable matchings, so hardness follows from Theorem~\ref{thm:unimodular}. Membership was discussed above; for bounded clique number the maximal cliques can be listed in polynomial time.
\end{proof}

Our orientations have no pair of opposite arcs, so hardness holds for orientations and hence also for the superorientations considered by Boros and Gurvich~\cite{BG96}.

\FloatBarrier
\section{Near-feasible stable matchings with couples}\label{sec:near}

Nguyen and Vohra~\cite{NV18} proved that every \HRC{} instance admits a stable matching after changing each hospital's capacity by at most two. Their construction first finds a fractional stable solution using Scarf's lemma and then rounds it. We show that finding the final near-feasible outcome is itself \PPAD-hard.

For a fixed allowance $r\ge1$, the problem \NFHRC{} asks for nonnegative integer capacities $\widehat c_h$ and a matching stable under them, where $|\widehat c_h-c_h|\le r$ for every hospital $h$. On the fixed-association instances of Corollary~\ref{cor:hrc}, a solution exists already with $\widehat c=c$.

\paragraph{The construction.}
Start with the reduction in Section~\ref{sec:uda} from a fan-out free \PC{} instance, and put $K=24r$. Replace every student by $K$ copies with the same preferences, and give every program quota $K$. In each program and university order, replace a student by her entire block of copies, ordered strictly within the block. Multiply university capacities by $K$, except that a \PUR{} university has capacity $13K/2$ instead of $7K$.

We use $D_i,T_i,R_{z,j}$ also for the corresponding blocks. For a node $z$, let $Z_j$ be the block replacing $z^j$, for $j=1,2,3$; for nodes $v,w$ we write $V_j,W_j$, respectively. Every member of $Z_j$ has the same first-choice home program $\home(z^j)$ and, if $z$ is read, the same second-choice input program $\inp(z^j)$. Each of these programs ranks $Z_j$ above its other applicants. The members of $Z_1\cup Z_2\cup Z_3$ are the $z$-students.

\paragraph{Correctness.}
Fix any allowed changes to all university capacities and program quotas, and a stable assignment $\mu$ under them. A program's new quota satisfies $\widehat q(p)\in[K-r,K+r]$, and a university of original capacity $C$ has new capacity in $[C-r,C+r]$.

We will repeatedly use the following consequence of stability. Suppose a block of $K$ students ranks a program $p$ first, the program ranks this block above all other applicants, and its university would accept every member of the block who is not at $p$. Then at least $\min(K,\widehat q(p))\ge K-r$ members of the block are at $p$.

Let $d(z)$ count the students in $Z_1\cup Z_2\cup Z_3$ who are away from their home programs, including unassigned students. Decode
\[
 x[z]=\begin{cases}
 0,&d(z)\le4r,\\
 1,&d(z)\ge3K-4r,\\
 \bot,&\text{otherwise}.
 \end{cases}
\]
For a gate $g$ reading $z$, let $a_g(z)$ be the number of $z$-students at their input programs. An unassigned input student cannot be rejected by the reading university: only other input students rank above her, and their number is at most $6K-1<9K-r$ in a \NOR{} gadget or $3K-1<13K/2-r$ in a \PUR{} gadget. Her input program must therefore be full. It has only $K$ applicants and quota at least $K-r$, so at most $r$ students of its block can be unassigned. Summing over the three blocks gives
\[
 d(z)-3r\le a_g(z)\le d(z).
\]
Therefore, a zero input implies at most $4r$ arrivals and a one input at least $3K-7r$. Write $a_g$ for the total number of arrivals to $g$.

\begin{lemma}\label{lem:near-nor}
Let $g=(\NOR,u,v,w)$ and $a=a_g$. If $a\le8r$, then $d(w)\ge3K-4r$. If $a\ge3K-7r$, then $d(w)\le4r$.
\end{lemma}
\begin{proof}
For $i=1,2,3$, let $t_i$ be the number of students of $T_i$ at $d_i$.

\emph{Case $a\le8r$.} University $U_g$ contains at most
\[
 a+7(K+r)\le7K+15r<9K-r
\]
students, so it is not full. Each $d_i$ therefore admits $\min(K,\widehat q(d_i))$ members of $D_i$, and $t_i\le r$ for $i\le3$.

If no $w$-student is at home, then $d(w)=3K$ and we are done. Otherwise $N_g$ contains a $w$-student, whom it ranks below every reservoir. It would therefore accept every reservoir. At least $K-t_i$ members of $T_i$ are outside $d_i$ and want $f_i$, unless already assigned there. Hence
\[
 |\mu(f_i)|\ge\min(K-t_i,\widehat q(f_i))\ge K-r.
\]
These programs occupy at least $3K-3r$ places at $N_g$. Its capacity is at most $3K+r$, leaving at most $4r$ students at home. Thus $d(w)\ge3K-4r$.

\emph{Case $a\ge3K-7r$.} We first show that at least $3K-3r$ reservoirs are at $U_g$. If $U_g$ contains no $D$-student, it must be full: otherwise a member of $D_7$ could take a free place at the empty program $d_7$ and block. Since at most $6K$ input students are there, at least $3K-r$ reservoirs are at $U_g$.

Otherwise $U_g$ contains a $D$-student and would accept every reservoir. Suppose $k\in\{1,2,3\}$ is minimal with $t_k<K-r$. A member of $T_k$ missing from $d_k$ must be rejected by that program. Thus $d_k$ is full and contains a $D_k$-student. For every $m>k$, the university ranks all members of $D_m$ above this student, and $d_m$ ranks $D_m$ first. Hence at least $K-r$ members of each such block are at $d_m$.

Count the earlier $k-1$ reservoir blocks, the full program $d_k$, and the $7-k$ higher-ranked $D$-blocks. Together with the arrivals, they give
\[
 |\mu(U_g)|\ge a+7(K-r)\ge10K-14r>9K+r,
\]
a contradiction. Therefore $t_i\ge K-r$ for every $i$, proving the reservoir bound in this case too.

At most $3r$ reservoirs can now be at $N_g$. If $N_g$ is full, at least $3K-4r$ of its students belong to the output blocks $W_1,W_2,W_3$. If it is not full, each home program contains at least $K-r$ members of its corresponding block $W_j$. In either case $d(w)\le4r$.
\end{proof}

\begin{lemma}\label{lem:near-pur}
Let $g=(\PUR,u,v,w)$ and $a=a_g$. If $a\le4r$, then $d(v),d(w)\le3r$; if $a\ge3K-7r$, then $d(v)=d(w)=3K$. For arbitrary $a$, either $d(w)\le3r$ or $d(v)=3K$.
\end{lemma}
\begin{proof}
\emph{Case $a\le4r$.} University $U_g$ contains at most
\[
 a+6(K+r)\le6K+10r<13K/2-r
\]
students and is not full. Each home program ranks its corresponding block $V_j$ or $W_j$ first. It therefore contains at least $K-r$ students from that block, giving $d(v),d(w)\le3r$.

\emph{Case $a\ge3K-7r$.} Suppose an output student is at home. The university ranks every reservoir above her, so it would accept every reservoir. Each of the four programs with a reservoir block must then contain at least $K-r$ students: either all $K$ reservoirs are there, or an absent reservoir forces the program to be full.
Moreover, at least one member of \(W_3\) is at home. Otherwise, since every member of \(W_3\) is ranked above any other output student, a member of \(W_3\) could block with her empty home program (the university has a worse output student by the assumption). As \(h(w^3)\) is not one of the four programs containing reservoir blocks, we obtain
\[
|\mu(U_g)|\ge a+4(K-r)+1
\ge 7K-11r+1 >
13K/2+r,
\]
a contradiction.

Thus, no output student is at home, so $d(v)=d(w)=3K$.

\emph{Arbitrary $a$.} If a $v$-student is at home, the university would accept every $w$-student, since all of them rank above her. Each home program of $w$ therefore contains at least $K-r$ members of its block $W_j$, giving $d(w)\le3r$. If no $v$-student is at home, then $d(v)=3K$.
\end{proof}

\begin{theorem}\label{thm:near}
For every fixed integer $r\ge1$, \NFHRC{} is \PPAD-hard, even in the fixed-association subclass, with at most two acceptable pairs per couple and capacities at most $216r$.
\end{theorem}
\begin{proof}
For a \NOR{} gate, two zero inputs give $a_g\le8r$, so Lemma~\ref{lem:near-nor} gives output one. An input one gives $a_g\ge3K-7r$, so the output is zero. Lemma~\ref{lem:near-pur} shows that a \PUR{} gate copies a pure input to both outputs and always has a pure output. Hence every stable assignment under any allowed capacity changes decodes to a \PC{} solution.

Interpret students as couples as in Corollary~\ref{cor:hrc}. Every couple lists at most two pairs, the university--program association is unchanged, and all capacities are at most $9K=216r$. A stable matching exists at the original capacities. For fixed $r$, construction and decoding are polynomial, proving the theorem.
\end{proof}
Nguyen and Vohra's guarantee also bounds the aggregate capacity change. Our reduction allows arbitrary individual changes, so the same proof applies under their precise bounds.

\begin{corollary}\label{cor:near-two}
\NFHRC{} is \PPAD-hard with allowance two even under the additional requirement $0\le\sum_h(\widehat c_h-c_h)\le4$, and with original capacities at most $432$.
\end{corollary}

\FloatBarrier
\section{Conclusion}\label{sec:conclusion}

We showed that integral stable solutions guaranteed by polyhedral integrality are \PPAD-hard to find. For unit capacities, hardness holds on unimodular hypergraphs that are linear and have hyperedges and degrees bounded by three, and therefore on balanced and on normal hypergraphs and for kernels in clique-acyclic orientations of perfect graphs. With capacities, hardness holds already on network hypergraphs with hyperedges of size three and capacities at most nine, in University Dual Admission, and in a natural subclass of dual markets with couples. The near-feasible extension shows that the same difficulty persists when each hospital may change its capacity by any fixed additive amount.

The results leave a clear map of the remaining open cases. Stable matchings are polynomial on network hypergraphs~\cite{CK26} and \PPAD-complete on unimodular ones. By Seymour's decomposition theorem~\cite{Sey80}, totally unimodular matrices are built from network matrices, their transposes, and a single exceptional matrix. Is \SHM{} polynomial on hypergraphs whose incidence matrix is the transpose of a network matrix? For kernels, our graphs are line graphs with clique number three; the complexity for claw-free perfect graphs with superorientations, left open in~\cite{PIM20}, remains unresolved. Further questions are the complexity of \SHbM{} on subpath hypergraphs~\cite{BCS25} and the smallest university capacity sufficient for \PPAD-hardness in \UDA. Finally, we hope that our techniques will be useful for showing hardness of other combinatorial problems where integral solutions are guaranteed.

\section*{Declaration of AI use}
During the research, the author used Claude (Anthropic) and ChatGPT (OpenAI) to assist in exploring proof strategies, developing and testing candidate gadget constructions for the reductions in Sections~\ref{sec:uda} and~\ref{sec:unimodular}, and searching the literature. The author contributed key mathematical ideas, proposed gadget outlines, directed the exploration of proof strategies, and critically evaluated and refined the AI-generated arguments and constructions. In particular, the author worked extensively on simplifying the reductions, streamlining the gadget constructions, and developing rigorous, intuitive, and accessible proofs.

The author checked all definitions, constructions, proofs, and references and takes full responsibility for the content of this paper. The AI tools are not authors of this work.

\end{document}